\documentclass{aastex61}

\usepackage{natbib}

\usepackage{xcolor}
\shorttitle{Genesis and impulsive evolution of the 2017 September 10 coronal mass ejection}

\shortauthors{}

\begin{document}

\title{Genesis and impulsive evolution of the 2017 September 10 coronal mass ejection}

\correspondingauthor{Astrid M. Veronig}
\email{astrid.veronig@uni-graz.at}

\author{Astrid M. Veronig}
\affiliation{Institute of Physics, University of Graz, Austria}
\affiliation{Kanzelh\"ohe Observatory of Solar and Environmental Research, University of Graz, Austria}

\author{Tatiana Podladchikova}
\affiliation{Skolkovo Institute of Science and Technology, Moscow, Russia}

\author{Karin Dissauer}
\affiliation{Institute of Physics, University of Graz, Austria}

\author{Manuela Temmer}
\affiliation{Institute of Physics, University of Graz, Austria}

\author{Daniel  B. Seaton}
\affiliation{Cooperative Institute for Research in Environmental Science, University of Colorado at Boulder, CO, USA}
\affiliation{National Centers for Environmental Information, National Oceanic and Atmospheric Administration, Boulder, CO, USA}

\author{David Long}
\affiliation{UCL-Mullard Space Science Laboratory, Holmbury St Mary, Dorking, Surrey, UK}

\author{Jingnan Guo}
\affiliation{Institut f\"ur Experimentelle und Angewandte Physik, University of Kiel, Germany}

\author{Bojan Vr\v{s}nak}
\affiliation{Hvar Observatory, Faculty of Geodesy, University of Zagreb, Croatia}

\author{Louise Harra}
\affiliation{UCL-Mullard Space Science Laboratory, Holmbury St Mary, Dorking, Surrey, UK}

\author{Bernhard Kliem}
\affiliation{Institute of Physics and Astronomy, University of Potsdam, Germany}

\begin{abstract}
The X8.2 event of 10 September 2017 provides unique observations to study the genesis, magnetic morphology and impulsive dynamics of a very fast CME. Combining GOES-16/SUVI and SDO/AIA EUV imagery, we identify a hot ($T\approx 10-15$ MK) bright rim around a quickly expanding cavity, embedded inside a much larger CME shell ($T\approx 1-2$ MK). The CME shell develops from a dense set of large AR loops 
($\gtrsim$0.5~$R_s$), and seamlessly evolves into the CME front observed in LASCO C2. 
The strong lateral overexpansion of the CME shell acts as a piston initiating the fast EUV wave. The hot cavity rim is demonstrated to be a manifestation of the dominantly poloidal flux and frozen-in plasma added to the rising flux rope by magnetic reconnection in the current sheet beneath.  The same structure is later observed as the core of the white light CME, challenging the traditional interpretation of the CME three-part morphology.  The large amount of added magnetic flux suggested by these observations 
explains the extreme accelerations of the radial and lateral expansion of the CME shell and cavity, all reaching values of 5--10~km~s$^{-2}$. The acceleration peaks occur simultaneously with the first RHESSI 100--300 keV hard X-ray burst of the associated flare, further underlining the importance of the reconnection process for the impulsive CME evolution. 
Finally, the much higher radial propagation speed of the flux rope in relation to the CME shell causes a distinct deformation of the white light CME front and shock.
\end{abstract}

\keywords{Sun: activity --- Sun: corona --- Sun: coronal mass ejections (CMEs) --- Sun: flares}

\section{Introduction}

Coronal mass ejections (CMEs) are large-scale structures of magnetized plasma that are expelled from the Sun with speeds ranging from some 100 up to about 3500~km s$^{-1}$ \citep{StCyr99,gopalswamy09c}, driven by magnetic forces \citep[e.g. reviews by][]{forbes06,chen11,green18}.  They are the most energetic events in our solar system, being associated with energy releases of up to some $10^{32}$ erg \citep[][]{vourlidas10,emslie12}.  Fast CMEs are often associated with EUV waves, which are widely believed to be low-coronal signatures of large-amplitude fast-mode magnetosonic waves or shocks 
\citep[e.g. reviews by][]{patsourakos12,liu14,warmuth15,long17}.
When the interplanetary counterpart of a CME, the ICME, and its associated shock reaches Earth, they  may induce strong geomagnetic storms, with the storm strength mainly depending on the southward ($B_z$) component of the ICME's magnetic field and on its impact speed \citep[][]{tsurutani88,tsurutani92,gonzalez94}.  
Recent studies, in particular owing to the multi-spacecraft in-situ observations of the STEREO satellites,
revealed the production of wide-spread solar energetic particles (SEPs), which 
seem to be able to fill the whole heliosphere  \citep{dresing12,gomez15,lario16}. It is understood that these extreme cases of wide-spread SEP events are either related to specific properties of the CME and its associated shock close to the Sun (extended source region) and/or to perpendicular particle transport processes in interplanetary space \citep[e.g.][]{dresing14}.
Thus, the understanding of the origin and early evolution of CMEs close to the Sun, their interplanetary propagation as well as their interaction processes with the Earth magnetosphere are all key aspects in the understanding and prediction of extreme space weather events \citep[e.g.][]{koskinen17}. 

CMEs observed in white-light coronagraphs often reveal a three-part structure: a leading bright front  followed by a dark cavity (void) and an embedded bright core \citep{illing86}. 
The bright front is a shell of enhanced density, due to compressed and piled-up material ahead of the erupting structure.
The dark cavity and  the bright core are generally interpreted as manifestions of a magnetic flux rope, i.e.\ a coherent helical magnetic structure with the field lines wrapping around the central axis \citep[e.g.,][]{low95}.  The flux rope 
may fill the whole volume of the dark coronal cavity, and thus the cavity may outline the flux rope's cross-section  in the plane-of-sky  (\citeauthor{chen97} \citeyear{chen97}; see also the review by \citeauthor{cheng17} \citeyear{cheng17}). 
The bright core is usually interpreted as corresponding to the cool and dense prominence material that is suspended in the magnetic dips at the bottom of the twisted flux rope \citep[e.g.,][]{Dere99,Gibson06}.
However, \cite{howard17} challenged this view, suggesting that the bright core represents the erupting flux rope with prominence material contributing to the emission only in a minority of events.
\cite{vourlidas13} pointed out that the CME front in the classical three-part structure may actually reveal a ``two-front" morphology. The bright loop-like front observed in the coronagraphs corresponds to the piled-up materal ahead of the erupting structure.  In addition, also a fainter outer front may be visible in the white-light coronagraph images (first reported in \cite{vourlidas03}, and reported in a number of events thereafter) as a result of the density enhancement by the shock that is driven by quickly accelerating CMEs. 

Recently, the early evolution of CMEs, their flux ropes and cavities in the inner corona have become accessible in the high-cadence multi-band  EUV observations from STEREO EUVI and SDO AIA. This is an important height range, where many CMEs appear to form and to undergo their impulsive acceleration phase \cite[][]{temmer10,bein11,patsourakos10b,patsourakos10,cheng11}. 
Several studies have shown that hot flux ropes may be enclosed in a dark cavity observed in cool EUV passbands,  showing plasma at typical quiet coronal temperatures \citep[][]{cheng11,cheng13,zheng12}.  A statistical study of eruptions observed in the EUV by \cite{nindos15} revealed that about half of the events exhibit a hot flux rope.

In general, CMEs observed in the coronagraph field-of-view (i.e.\ beyond several solar radii) reveal a self-similar expansion, i.e.\ the radial and lateral expansion develop at the same rate \citep{schwenn05}. Recent studies of  CME and cavity evolution in high-cadence EUV imagery report fast lateral overexpansion of CMEs low in the corona, i.e. the radius (width) of the CME expands stronger than it gains in height \cite[][]{patsourakos10b,patsourakos10,cheng13}. Further, fast lateral CME expansion has been shown to be important for the formation of EUV waves and shocks low in the corona \citep[e.g.][]{veronig08,temmer09,patsourakos10b,veronig10,cheng12}.

In this paper, we present a case study of the origin and early evolution of the fast CME that occurred together with the X8.2 flare on 10 September 2017, and its associated EUV wave.  
The event was the second largest flare in solar cycle 24, and extreme in several aspects.
It was associated with a very fast halo CME with a speed $\gtrsim$ 3500~km~s$^{-1}$ \citep[][]{guo18,gopalswamy18}, and  revealed distinct signatures of a long hot current sheet that formed behind the erupting structure
\citep{seaton18,yan18,warren18}. The event produced wide-spread SEPs detected at Earth, Mars and STEREO-A, i.e.\ covering a width of at least 230$^\circ$ in helio-longitude. Notably, this SEP event was the first Ground Level Enhancement (GLE) that was observed on the surface of two planets, Earth and Mars \citep[][]{guo18}.
Fermi-LAT observed an extremely long duration $\gamma$-ray event, detecting $>$100 MeV emission from the flare that lasted for more than 12 hours \citep{omodei18}. The associated EUV wave was also unique, as it was globally propagating across the full solar disk as observed from Earth view as well as from STEREO-A, and showed transmission through both polar coronal holes (\citeauthor{liu18} \citeyear{liu18}; Podladchikova et al., in preparation).
Some further aspects of this intriguing event that have been studied are the formation and turbulent plasma motions in the current sheet \citep{Li18,warren18}, the evolution of the flux rope, its cavity and reconnection signatures \citep{seaton18,yan18,long18}, the coronal plasma properties in the aftermath of the CME \citep{goryaev18}, the microwave and hard X-ray flare emission \citep{gary18}, 
 the shock properties near the Sun and the characteristics of the shock-accelerated particles 
\citep{gopalswamy18} as well as the the coronal-to-interplanetary CME evolution 
and associated wide shock \citep{guo18}.
 
We concentrate on the genesis, morphology and impulsive dynamics of the CME in the low-to-mid corona, as well as on its relation to the formation of the associated fast EUV wave. 
Very valuable observations on these phenomena are provided by 
the new Solar Ultraviolet Imager \cite[SUVI;][]{seaton18} on board GOES-16.  SUVI observes the solar EUV corona over a large field-of-view (FOV), allowing us to robustly connect CME structures in subsequent images without change of instrument/emission processes, and with a cadence high enough to resolve the dynamical processes during the impulsive CME phase.
These data are complemented by the high-cadence multi-band EUV imagery of the Atmospheric Imaging Assembly \cite[AIA;][]{lemen12} onboard the Solar Dynamics Observatory \cite[SDO;][]{pesnell12}, to study the impulsive evolution of the CME cavity and flux rope low in the corona. As we will demonstrate in this paper, the event under study provides us also with unique observations of the three-part CME structure from the EUV to the white-light coronagraph data, and new insights into its physical interpretation.

\section{Data}

The SUVI instrument on board of the GOES-16 spacecraft images the solar corona in six EUV passbands, centered at 94, 131, 171, 195, 284, and 304 {\AA}, with a large FOV, out to $>$1.6~$R_s$ in the horizontal direction and as large as 2.3~$R_s$ in the unvignetted corners \citep{seaton18}, with $R_s$ denoting the solar radius. The pixel scale of the SUVI filtergrams is 2.5~arcsec. SUVI's camera is equipped with an anti-blooming circuitry, which enables unobscured observations of the flare regions even in exposures where the detecor is substantially saturated, while at the same time providing high sensitivity to observe faint structures (e.g., EUV waves). The AIA instrument onboard SDO observes the solar EUV corona in seven wavelength bands, centered at 94, 131, 171, 193, 211, 335, and 304 {\AA}, with a cadence as high as 12 s, a pixel scale of 0.6~arcsec and over a FOV of about 1.3~$R_s$ in horizontal direction. Thus, AIA and SUVI both sample solar plasmas over a broad temperature range of about $10^5$ to $10^7$~K.

The X8.2 event of 10 September 2017 occurred when the source active region NOAA 12673 was on the Western limb (flare start/peak time: 15:35/16:08 UT; heliographic coordinates: S08W88).
SUVI and AIA observed the flare, the CME initiation and evolution in the low-to-mid corona as well as the associated EUV wave.
We use the SUVI 195 {\AA} filter observations available with a mean cadence of 1~min to study the impulsive radial and lateral expansion of the CME shell as well as the associated EUV wave. 
To enhance the structures in the images, in particular above the limb, we use the same filtering technique as in \cite{seaton18}. This is a two-step filtering consisting of a varying radial filter (to account for the coronal intensity fall off in the corona) and a temporal unsharp masking technique (to enhance changes in time). The image series was compensated for solar differential rotation, and base difference images were created.  
In addition, we use the multi-wavelength SUVI imagery to study the radial kinematics of the flux rope and cavity \citep{seaton18}.
Note that we did not rotate the SUVI images to North up, in order to keep the full FOV that SUVI provides; the solar P0 angle of the day was $+23.1^\circ$.
 
The SUVI data are complemented by the multi-temperature EUV imagery from the AIA instrument onboard SDO, to study the radial and  lateral expansion of the distinct cavity observed low in the corona \citep[][]{long18}. Each AIA image has been processed using the Multi-scale Gaussian Normalization technique of \cite{morgan14} to highlight the fine structures above the limb. Data from the Large Angle and Spectrometric Coronagraph 
\citep[LASCO;][]{brueckner95} onboard the Solar and Heliospheric Observatory (SOHO) are used to set the CME shell and cavity observed in the EUV into context with the white-light CME observed further out by the LASCO C2 and C3  coronagraphs.
The time line of the associated flare is studied in the GOES soft X-ray (SXR) and  Ramaty High Energy Solar Spectroscopic Imager \cite[RHESSI;][]{lin02} hard X-ray (HXR) data.

\begin{figure}
   \centering
	\includegraphics[width=0.65\textwidth]{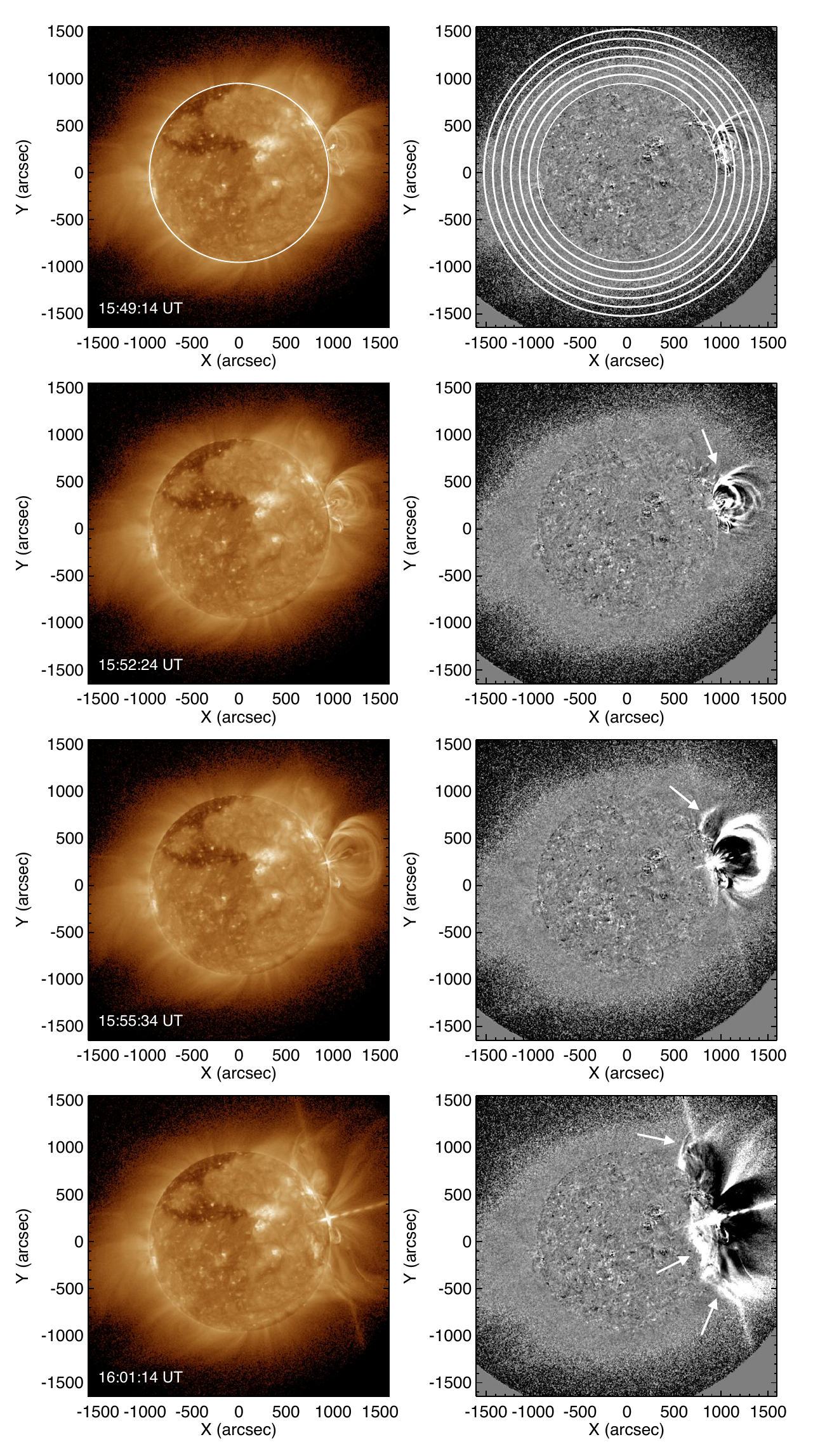}
	\caption{Overview of the 10 September 2017 event in SUVI 195 {\AA} filtergrams. Left: direct images, right: base difference images. In the first difference image, we overplot circles at heights of 1.0 to 1.6~$R_s$ in steps of 0.1~$R_s$ used to derive the stack plots shown in Figure~\ref{fig:suvi_stacks}. The arrows indicate the EUV wave observed above the limb and on the disk. The still figure shows the event at four time steps (annotated in each panel). The animated figure online shows the event evolution from 15:20:24 to 17:00:24 UT.	An animation of this figure is available in the online journal. }
	\label{fig_overview}
\end{figure}

\begin{figure}
\centering
	\includegraphics[width=0.85\textwidth]{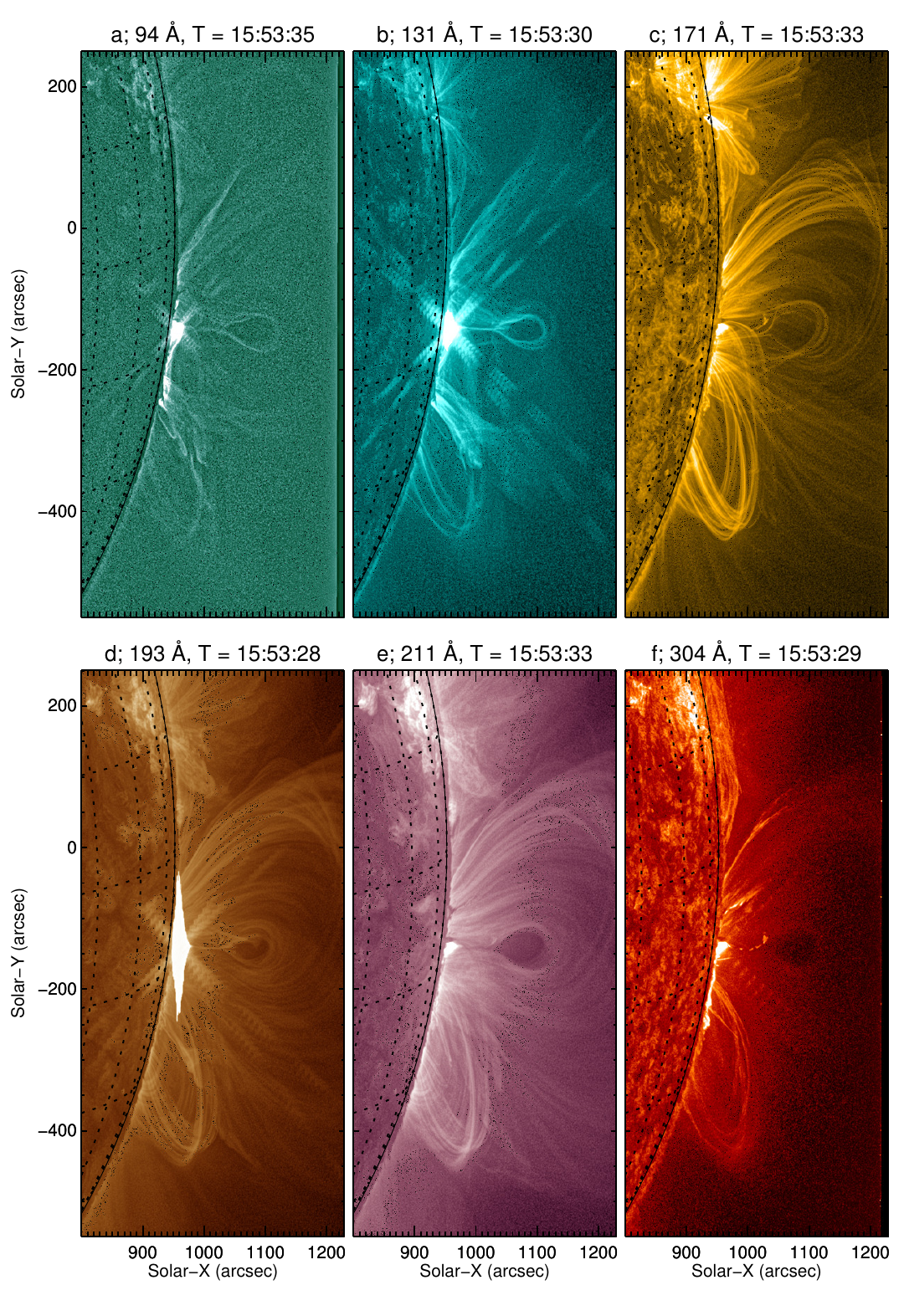}
	\caption{
	CME cavity  as observed by SDO/AIA in the 94, 131, 171, 193,  211, and 304 {\AA} filters. The still figure shows the cavity at one time step for each filter (annotated at each panel). The animated figure online shows the cavity evolution from about 15:17:30 to 16:40:00 UT.
	An animation of this figure is available in the online journal. 
}
	\label{fig_cavity}
\end{figure}

\section{Results}

\subsection{Event overview}
\label{overview}

Figure~\ref{fig_overview} and the accompanying movie give an overview of the early CME evolution and the formation of the EUV wave as observed in SUVI 195 {\AA} direct images and corresponding base difference images. In the movie, we can clearly see the expansion of the large pre-existing loop system that  develops into the CME shell as well as the development of the associated EUV wave. We note that the pre-eruptive structure of the source AR 12673 is very distinct, consisting of a dense set of very large loops, with sizes up to $>$0.5 $R_s$. These are a result of the  extremely strong fields of $>$5000 G that have been measured in AR 12673 \citep{wang18}.
In the movie, it is seen that the first motions and re-configurations related to the event can be observed as early as 15:39 UT, including the rise of a low-lying, already brightened prominence. During its rise, the prominence  appears to dissolve and the eruption develops an inner flux rope structure with a clear cavity that is quickly expanding \citep[see also][]{long18,seaton18,yan18}.

More detail of the CME formation and its early evolution can be seen in Figure~\ref{fig_cavity} and the associated movie, which shows a zoom to the  eruption region in six of the AIA EUV passbands. In the hot AIA  94 {\AA} channel (dominated by emission from an Fe {\sc xviii} line; peak formation temperature $T\approx 7$ MK), the cavity is observed earliest and also some embedded prominence material can be observed. The cavity becomes more pronounced during its fast rise and expansion. It appears dark in all AIA channels, indicative of a strong density depletion inside the cavity due to the enhanced magnetic pressure of the enclosed flux rope \citep[e.g.][]{gibson15}, and is surrounded by a distinct bright rim that can be seen in all AIA EUV passbands.
Throughout the eruption, the bright rim is most distinct in the AIA~131 {\AA} filter (Fe {\sc viii} and {\sc xxi}), sampling plasmas at temperatures $\gtrsim 10$ MK.
This suggests that the bright rim around the cavity may be a manifestation of the magnetic flux and frozen-in plasma that is fed to the erupting flux rope by the magnetic reconnection in the current sheet beneath, and that, hence, the hot cavity rim outlines the cross-section of the enclosed expanding flux rope (as 
the simulations by \cite{lin04} suggest). This interpretation is further supported by the thin elongated structure that connects the bottom of the hot rim around the cavity to the top of the rising flare loop system, most clearly observed in the AIA~131 {\AA} filter (see Figure~\ref{fig_cavity} and associated movie). This structure has been interpreted as an observational signature of the hot plasma around the current sheet that is formed behind the eruption \citep{seaton18,yan18,warren18}.
At later stages ($\gtrsim$16:00 UT) after the cavity has already exited the AIA  FOV, the elongated current sheet appears well pronounced in all the AIA EUV channels. As shown in \cite{warren18}, the temperature in the current sheet is about 20 MK, with a relatively narrow distribution. The observed sheet thickness is  about 3000 km \citep{yan18}. 

At the very start of the event, during 15:39--15:51 UT, hot loops are seen to reconnect and to form new configurations, as revealed in the AIA  131 and  94 {\AA} images in Figure \ref{fig_cavity} and the associated movie. We can see changes in different flux systems connecting to the top of the flux rope 
structure.  These changes are observed immediately before the cavity and bright rim are clearly formed, and seem to be the trigger for the fast eruption.

The EUV wave associated with this event is distinctly seen above the solar limb as well on the solar disk (Figure~\ref{fig_overview} and associated movie; see also \cite{seaton18,liu18}). 
In the SUVI frame at 15:52:24 UT, we can identify the wave for the first time, as it is formed ahead of the CME flanks expanding toward the North (indicated by an arrow in Figure~\ref{fig_overview}). 
In this Northern direction, the EUV wave appears as a sharp front above the limb growing to a large extension in height, up to the borders of the SUVI field-of-view (FOV). 
Toward the South, the formation and detachment of the wave from the CME flanks is not as distinct as toward the North, but after about 15:58 UT also here the EUV wave can be observed ahead of the CME shell in its propagation above the limb. 
On the disk, the EUV wave can  be first identified in the SUVI frame at 15:54:24 UT. 
Surprisingly, despite the strength and global propagation of this EUV wave, inspection of GONG H$\alpha$ image series did not reveal any signature of an associated Moreton wave. 


The base difference images in Figure~\ref{fig_overview} reveal a large coronal dimming above the limb, consisting of different parts signifying different magnetic domains and physical phenomena. The main dimming is associated with the CME eruption, as it is laterally well confined to within the expanding CME bubble. This is the coronal dimming due to the expansion and evacuation of mass by the erupting CME \citep[e.g.][]{hudson97,zarro99,dissauer18}. 
The other dimming regions above the limb, further apart from the eruption center both in the North and South directions, are more shallow and are trailing the propagating EUV wave front. These findings suggest that they are signatures of the rarefaction region that forms behind the compression front of the wave \citep{muhr11,lulic13,vrsnak16}. 
The movie also reveals a number of interactions of the EUV wave with plasma structures in the corona, such as reflections, refractions and most interestingly also the transmission through the polar coronal holes.
These interactions and wave kinematics are studied in \cite{liu18} and  Podladchikova et al. (2018, in preparation).

\subsection{Early dynamics of the eruption} 

The event of 10 September 2017 provides us with an excellent opportunity to study the early impulsive dynamics of both the CME shell and the embedded flux rope/cavity in a very fast eruption. This is owed to the clearly formed  and well separated structures in this event captured by the high-cadence and large FOV observations of the AIA and SUVI EUV imagers, respectively. Both phenomena are important to study, to better understand the origin of the eruption and its consequences. The flux rope represents a current carrying structure that provides the forces driving the rise and expansion of the CME bubble. The dark EUV cavity that we observe is interpreted to map the flux rope's cross-section, as is further discussed in Sect.~\ref{sect_morphology}. The CME shell represents the outermost layer of compressed and piled-up material moving through the corona and interplanetary space. It also acts as the contact surface that creates shock waves ahead of it.

\begin{figure}
   \centering
	\includegraphics[width=0.9\textwidth]{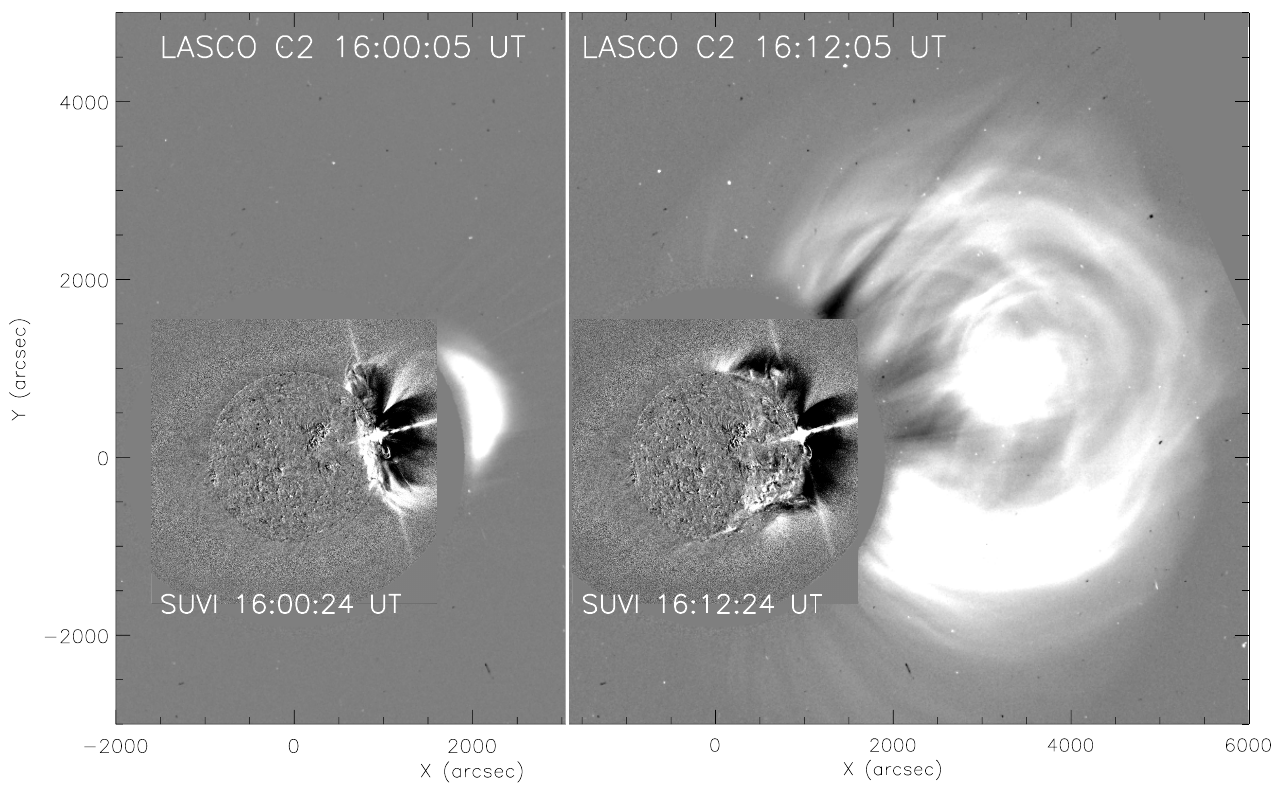}
	\caption{
	Composite SUVI 195 {\AA}  and LASCO C2 difference images, showing the connection of the CME outer front observed in the EUV 
	to the white-light-coronagraph data.  Note that the LASCO images are rotated to the SUVI orientation, i.e.\ to the P0 angle of $+23.1^\circ$.
	}
	\label{fig_lasco}
\end{figure}

Figure~\ref{fig_overview} and the associated movie show that following the early rise of the flux rope/cavity, the large pre-existing set of loops observed in the SUVI 195 {\AA} filtergrams also start to rise and to expand by successively piling up higher overlying sets of loops.  It is also seen that the lateral expansion of this forming CME shell is more pronounced than the radial one, leading to a distinct lateral overexpansion. 
In Figure \ref{fig_lasco}, we show composite images from the SOHO/LASCO C2 coronagraph  and co-temporal SUVI 195 {\AA} filtergrams for the first two time steps where the CME entered the C2 FOV.
The composite at 16:00 UT clearly shows that the outer boundary of the loop structures that we see expanding in the SUVI 195 {\AA} images seamlessly fits to the CME front observed in the white-light coronagraph data. In the Northern direction, also some fainter outer emission region in C2 is seen, which appears to connect to the EUV wave front observed in SUVI that is already well separated from the CME shell at that time. This indicates that this faint outer emission region in C2 is an early signature of the CME shock wave observed in white-light \cite[cf.,][]{vourlidas03}. Finally, in the C2 image at 16:12 UT a bright core is seen, indicative of the embedded flux rope.

\subsubsection{Early dynamics of the CME shell}
\label{sect_shell}
 
\begin{figure}
\centering
	\includegraphics[width=1.0\textwidth]{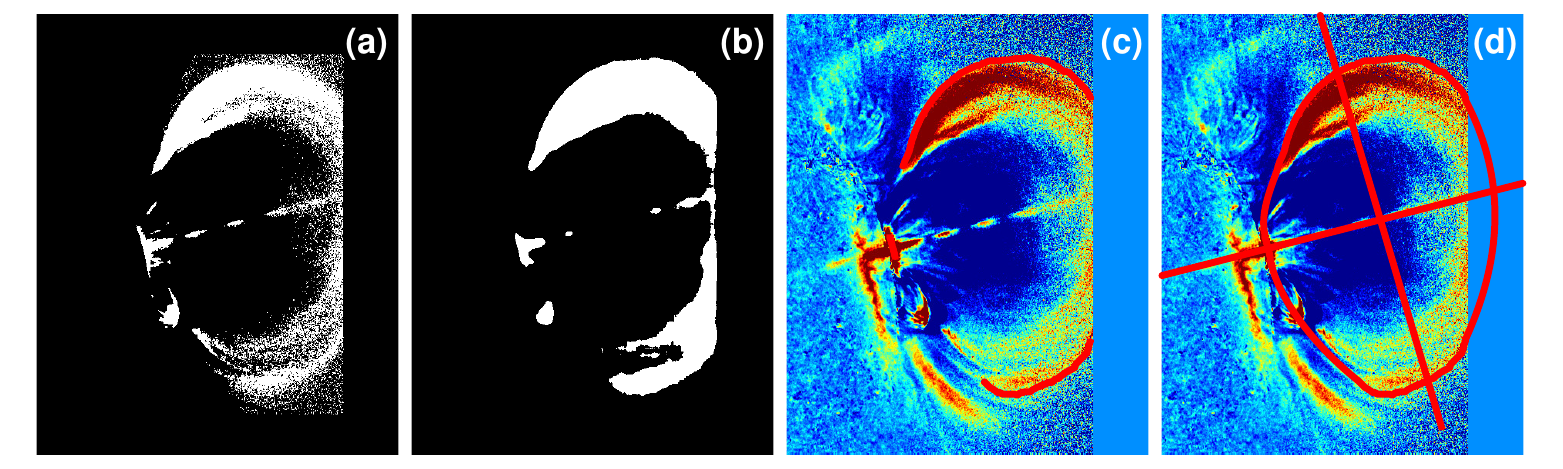}
	\caption{Illustration of the steps applied to segment the CME structure, shown for the SUVI 195 {\AA}
	base difference image at 15:57 UT. a) Binary map derived by thresholding. 
b) Segmented CME shell after median filtering. c) Borders of the visible parts of the CME (red contour) on top of the SUVI base difference image. d) CME borders including the interpolated parts (red contour). The straight red lines indicate the radial and lateral directions through the center of the CME. }
	\label{fig1}
\end{figure}

\begin{figure}
\centering
	\includegraphics[width=0.86\textwidth]{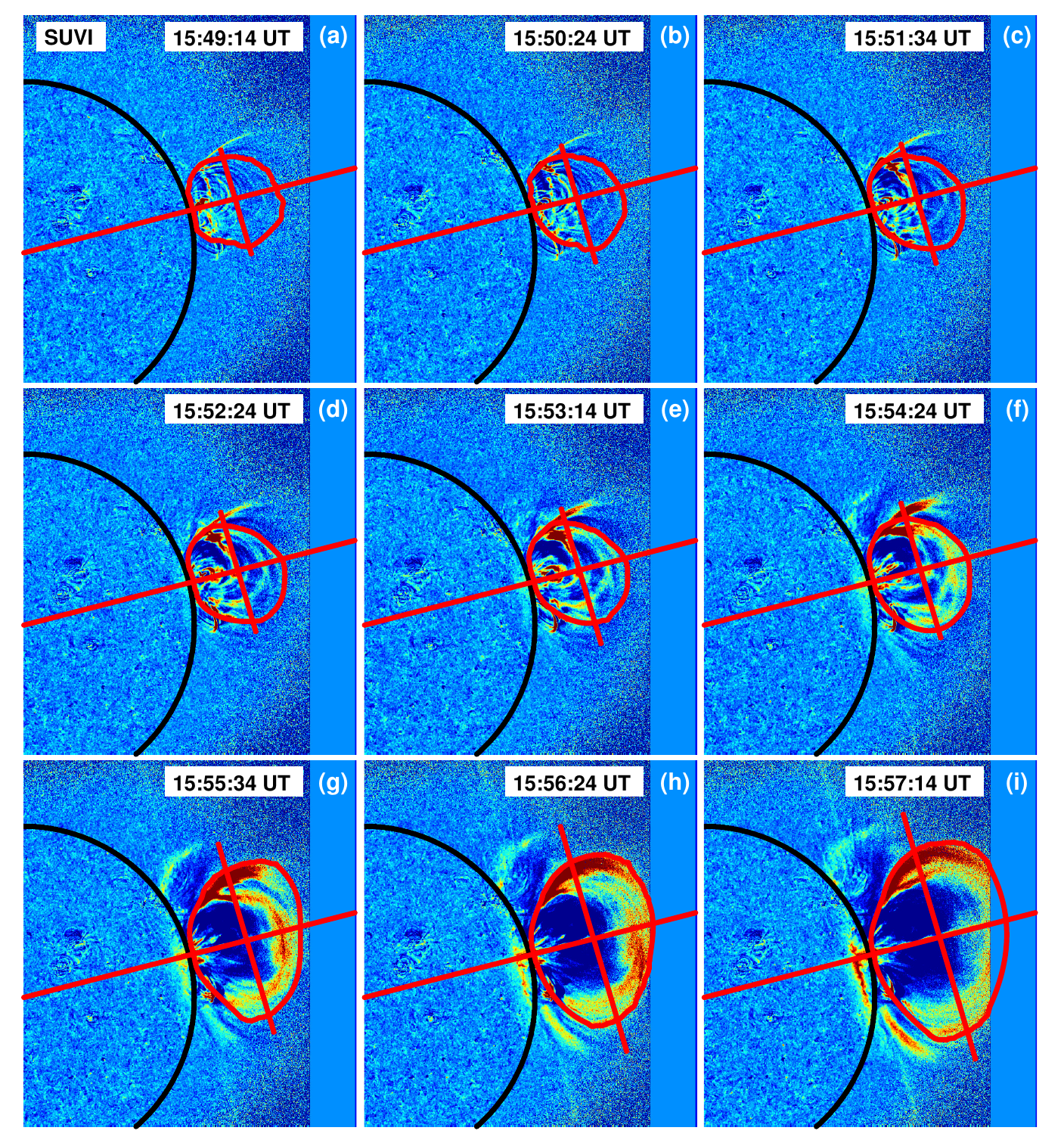}
	\caption{SUVI 195 {\AA} base difference images together with the outer borders of the segmented CME structure (red contours). The tracing of the CME radial and lateral expansion is indicated by the orthogonal straight red lines, respectively. }
	\label{fig2}
\end{figure}

In order to quantify the impulsive radial and lateral expansion of the CME shell in the SUVI FOV, we segment its structure and follow its evolution. At each time step, a binary map is constructed by thresholding, using the 2\% brightest pixels in the SUVI 195 {\AA} base difference images. 
Figure~\ref{fig1}a shows an example binary map derived, where all the white pixels are assumed to belong to the CME bubble. However, this segmentation includes also pixels without connection to the CME bubble, 
and these are subsequently removed by median-filtering (Figure~\ref{fig1}b). From this segmentation, we then determine the outer border of the CME structure and its $(x,y)$ coordinates as shown by the red contour in Figure~\ref{fig1}c. Finally, missing parts of the border are determined by interpolation of the  $(x,y)$ coordinates on the basis of minimization of the second derivative of the curve describing the border (Figure~\ref{fig1}d).  
Note that with this approach we are able to automatically detect the outer envelope of the set of expanding EUV loops that is subsequently developing into the white-light CME (cf.\ Figure~\ref{fig_lasco}). Applying the segmentations on the SUVI base difference images implies that we identify those regions which are increasing in emission due to expansion and subsequent pile-up of overlying loops, and thus becoming part of the overall erupting structure.

Figure~\ref{fig2} shows the results of this segmentation procedure for all SUVI 195 {\AA} images during 15:49--15:57 UT.  
As can be seen from the determined borders, the shape of the CME structure transforms from roundish to an ellipse, indicative of a strong lateral overexpansion during its evolution.  
We also note that the segmentation algorithm correctly handles the loop system along the northern boundary, which is observed  as a sharp, slightly-curved radial structure. It exists already before the event, and the algorithm correctly identifies it as not being part of the CME shell. However, when the pre-existing loop structure expands to form the CME, also the overlying loops are piled up and eventually become part of the CME shell. For the loop structure toward the North, this change is correctly identified between time steps 15:54 UT (where it is identified as an external structure, i.e.\ not part of the CME) and 15:55 UT (where it became integrated to the expanding CME shell). 

Based on the CME segmentations at each time step, we trace its outward propagation along the radial direction from Sun center to the center of the CME structure. The height of the CME is then determined as its radial extent plus the distance between its lower border and the solar surface. The CME lateral expansion is analyzed by constructing the perpendicular line through the center of the segmented structure and determining the intersection points with its borders to determine the CME width (cf.\ the straight lines in Figure \ref{fig2}).  
The relative errors on the derived height and width data are estimated to be about 2\%.
These height and width measurements are then used to study the kinematics and dynamics of the CME shell.
To obtain robust estimates of the corresponding velocity and acceleration profiles, we first smooth the height-time (width-time) curves, and then derive the first and second time derivatives.  
The smoothing algorithm that we use for approximating the curves is based on the method described in \cite{podladchikova17}, extended toward non-equidistant data. The 
algorithm optimizes between two (intrinsically conflicting) criteria in order to find a balance between data fidelity, i.e.\ the closeness of the approximating curve to the data, and smoothness of the approximating curve. The data fidelity is evaluated by minimizing the sum of the squared deviations between the fit and the data points, and the smoothness is evaluated by minimizing the sum of squared second derivatives of the fit curve.

\begin{figure}
\centering
\includegraphics[width=0.48\textwidth]{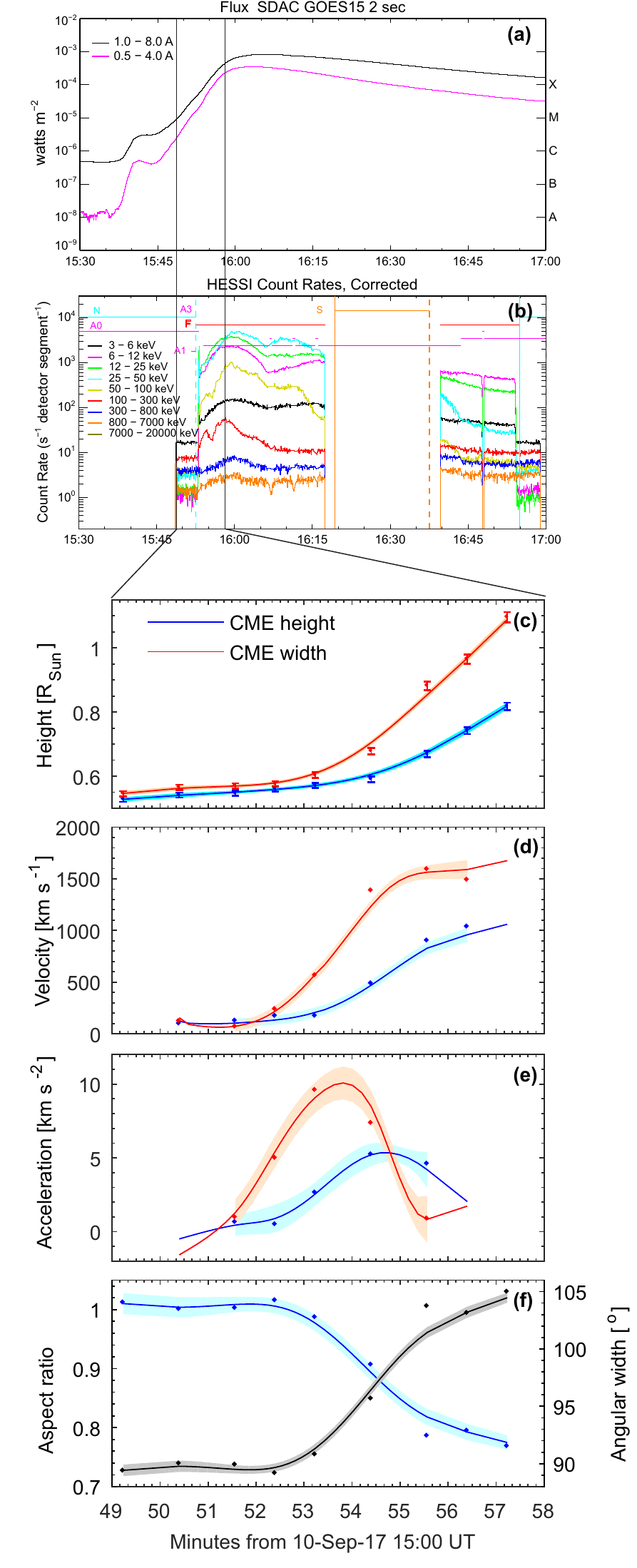}
	\caption{Impulsive phase of the radial and lateral evolution of the CME shell, and associated flare emissions. (a)
	Flare emission recorded in the GOES 0.5--4 and 1--8 {\AA} SXR bands, (b) RHESSI hard X-ray count rates in four energy bands from 12 to 300 keV. 
	(c) CME height (blue dots) and CME width (red dots) determined from SUVI 195 {\AA} images (cf.\ Figure \ref{fig2}) together with error bars. The corresponding lines show the smoothed height-time and width-time profiles. 
	 (d) Velocity, (e) acceleration of the CME radial (red) and lateral (blue) expansion obtained by numerical differentiation of the data points (dots) and the smoothed curves (lines). The shaded regions outline the error ranges obtained from the smoothed curves.
(f) CME aspect ratio (black) and angular width (blue). }
	\label{fig3}
\end{figure}

Figure~\ref{fig3} shows the evolution of the radial  and lateral expansion of the CME. 
In panel c, we plot the CME height (blue) and width (red) as function of time. 
Panels d and e show the velocity and acceleration of the CME in radial (height) and lateral (width) direction. 
In addition to the smoothed curves, we also show the velocity and acceleration curves derived by numerical differentiation of the data points themselves. As one can see, the numerical derivatives, which are intrinscially very sensitive to noise, are in general well aligned with the smoothed curves. Due to the fast evolution, the changes over the mean observing cadence of 1 min are big enough, so that in this case the noise on the direct derivatives is small. We also note that the slight ``jump"  in the CME lateral width evolution between 15:54 and 15:55 UT, which is due to the pile-up of the loop structure observed to the Northern boundary (as discussed above), is well handled by the smoothing algorithm applied on the kinematic curves (cf.\ Figure~\ref{fig3}c, red line). 
In panel f, we plot the evolution of the CME aspect ratio $r$, defined as the height of the center of the CME shell above the solar limb divided by its half-width \citep[cf.][]{patsourakos10b,patsourakos10}.  The aspect ratio is basically a measure of the opening angle (angular width) of the CME, in the present case defined from its source region on the solar limb. In panel f, we also plot the corresponding angular width of the CME derived as $W = 2 \arctan(1/r)$.
On the top panels, we plot (a) the GOES 1--8 {\AA} SXR flux  and (b) the RHESSI HXR count rates in four energy bands from 12 to 300 keV, indicative of the evolution of the energy release and particle acceleration in the associated flare.

As can be seen from the kinematical curves in Figure~\ref{fig3}, the CME reveals a very fast evolution. The radial propagation speed is continuously increasing, and reaches  $\sim$1000~km~s$^{-1}$ at a distance of 
0.85~$R_s$ above the limb.  
Later, in the coronagraphic FOV, the CME reached a maximum speed of  
$\gtrsim$3500~km~s$^{-1}$ \citep {guo18,gopalswamy18}. 
The lateral CME expansion reveals a speed of $\sim$1600~km~s$^{-1}$ in the SUVI FOV, reached within 4~min of impulsive acceleration. 
The acceleration curves reach their maxima both within the SUVI FOV, notably with much higher values for the lateral expansion, $10.1\pm 1.1$~km~s$^{-2}$ at 15:53:50 UT, while the radial peak acceleration is $5.3\pm 0.6$~km~s$^{-2}$ at 15:54:40 UT. The difference in the radial and lateral evolution causes a fast decrease of the CME aspect ratio, which changes from 0.96 to 0.75 within 5~min over a height range of just 0.2~$R_s$,  
corresponding to an increase of the CME angular width from about 90 to $105^{\circ}$.
We note that the peaks in the lateral and radial CME acceleration occur simultaneously within the SUVI measurement cadence of 1 min, and also simultaneous with the first RHESSI 100--300 keV HXR burst (15:54:40 UT). 

\subsubsection{Cavity morphology  and flux rope}
\label{sect_morphology}

As discussed in Sect. \ref{overview}, the cavity embedded inside the CME shell appears dark in all AIA EUV channels, suggestive of a strong density depletion due to the enhanced magnetic pressure in the enclosed magnetic flux rope, and it is surrounded by a bright rim (cf.\ Figure \ref{fig_cavity}). Throughout the eruption observed in the AIA FOV, this rim appears most pronounced in the 131 {\AA} filter, together with the hot   current sheet \cite[$T\approx  20$~MK;][]{warren18} beneath which connects the hot cavity rim to the top of the rising flare loop system. The high plasma temperatures in the bright rim around the expanding cavity are confirmed by the AIA Differential Emission Measure (DEM) analysis in \cite{yan18} and from Hinode/EIS spectroscopy, which shows the rim around the expanding cavity most prominently in the Fe {\sc xxiv} spectral line \citep[see Figure 4 in][]{long18}, sampling plasmas at temperatures of about 15~MK. Note that the DEM maps shown in \cite[][Figure 4 therein]{yan18} suggest that the thin cavity plasma embedded {\em inside} the bright hot rim is also hot.

\begin{figure}
\centering
\includegraphics[width=0.99\textwidth]{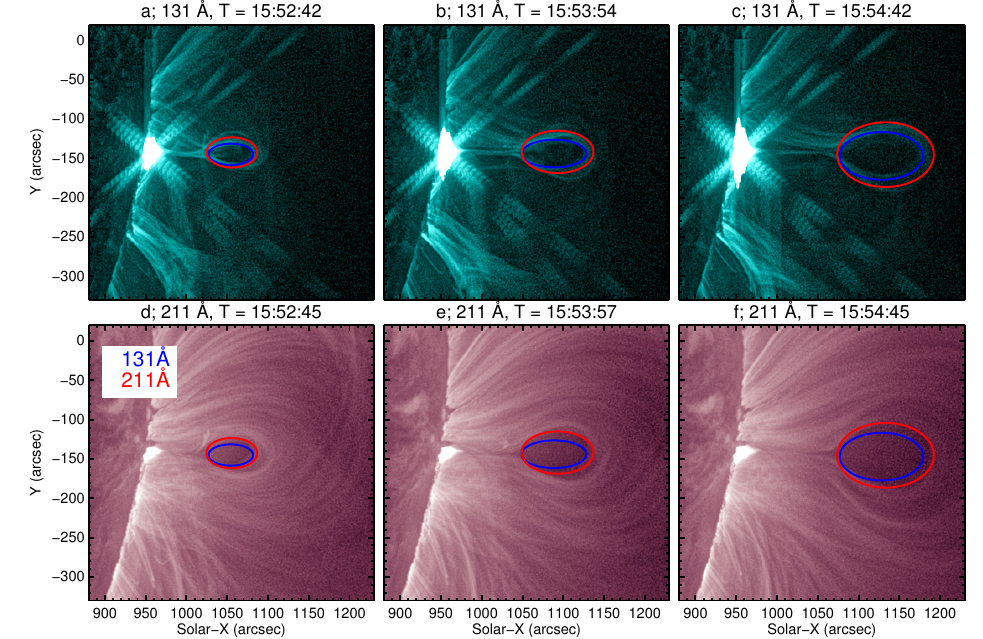}
	\caption{
	Snapshots of the eruption observed in the AIA 131 {\AA} (top) and 211 {\AA} (bottom) filters. The blue (red) contours outline the ellipses fitted to the inner edge of the bright rim around the dark cavity as observed in AIA 131 (211) {\AA}. The ellipse fits are from \cite{long18}.
	}
	\label{fig_ellipse}
\end{figure}

It is important to  note that a bright rim around  the cavity is observed in all AIA channels, i.e.\ those sampling hot plasma at about 10 MK (131, 94 {\AA}) as well as the ``cooler" filters which are  most sensitive to  plasma at typical quiet coronal temperatures in the range 1--2 MK (171, 193, 211 {\AA}). However, there are also differences  in the cavity/rim morphology as observed in the hot and cooler filters, which have important implications. 
In particular, we  note that the bright rim (and enclosed cavity) have different sizes in the different AIA filters, appearing smallest in the hot AIA passbands. This size difference is clearly seen in the co-temporal AIA 131 and 211 {\AA} snapshots plotted in Figure \ref{fig_ellipse}, where the hot bright rim observed in AIA 131 {\AA} well fits inside the dark cavity observed in the AIA 211 {\AA} filter.  

The smaller hot rim around the cavity, as most distinctly observed in the 131 {\AA} filter, is strongly suggestive of additional magnetic flux that is supplied to the flux rope by magnetic reconnection in the current sheet beneath. Since the plasma contained in this flux is heated by the reconnection as it passes through the current sheet, it does not contribute to the emission in the EUV channels sensitive to lower temperatures, so that these passbands show the whole cross-section of the flux rope as a dark cavity. 
These observations provide strong evidence that the distinct EUV cavity and surrounding hot rim observed in this event outline the cross-section of the expanding flux rope including the layer of newly added  flux by magnetic reconnection in the current sheet.

The bright rim observed in cooler plasmas like in 211  {\AA}, which has its inner boundary located at the outer boundary of the hot rim observed in 131  {\AA} (red ellipses in Figure \ref{fig_ellipse}; see also the movie associated with Figure \ref{fig_cavity}), is most likely formed by the compression of the plasma in the nearby overlying loops that get piled-up by the rising and expanding flux rope. This is before the corresponding field  lines also come to magnetic reconnection in the current sheet beneath, and a hot rim is formed around the cavity where a cool rim was observed before.

\begin{figure}
\centering
\includegraphics[width=0.52\textwidth]{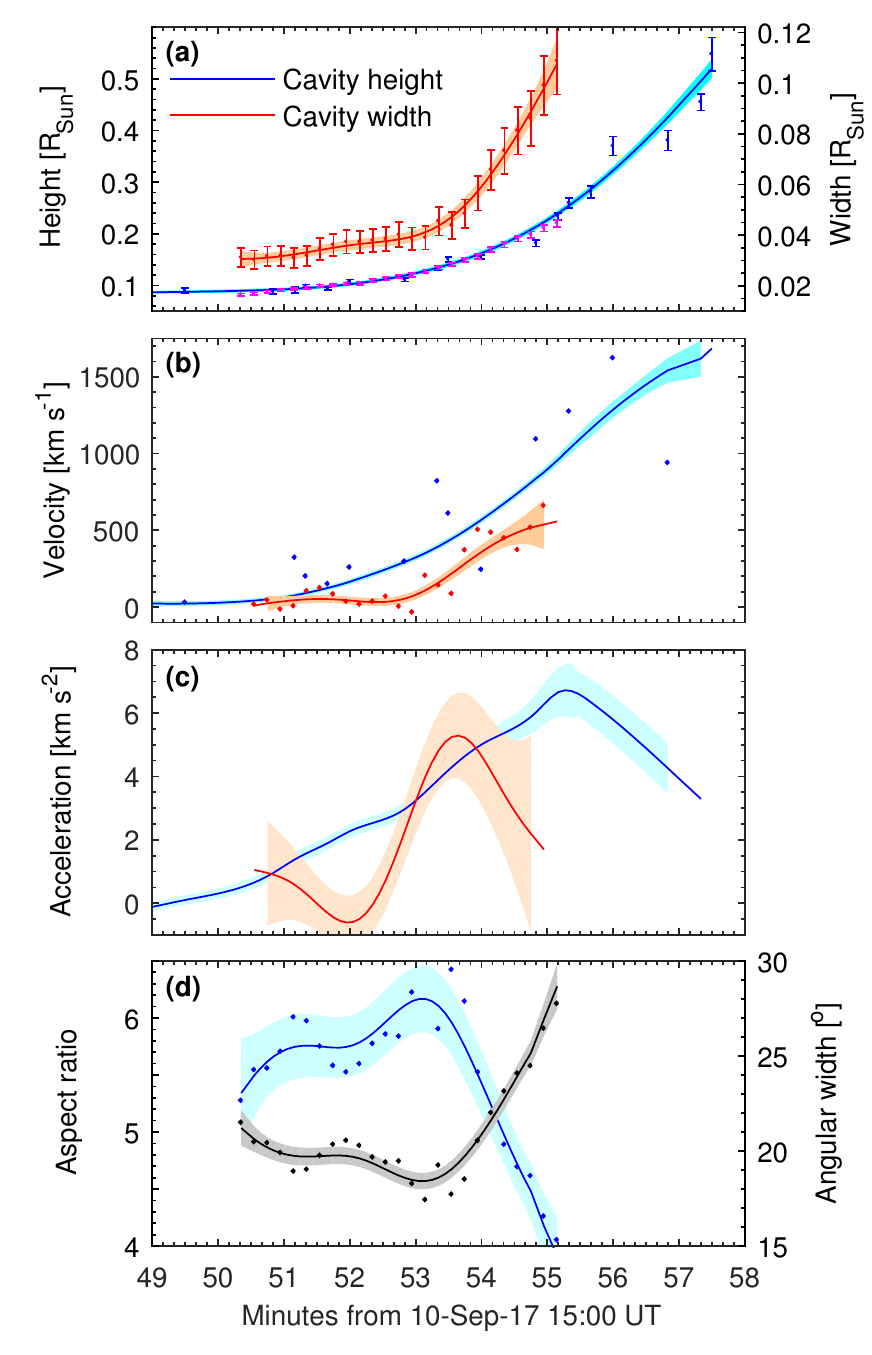}
	\caption{
	Evolution of the CME cavity. 
	(a) Height and width of the CME cavity, together with error bars. The corresponding lines show the smoothed height-time and width-time profiles.	Height is determined from multi-wavelengths SUVI data (blue; measurements from \cite{seaton18}) and AIA 211  {\AA} images (pink; measurements from \cite{long18}). Width is determined from  AIA 211 {\AA} images (red; measurements from \cite{long18}).
	 (b) Velocity, (c) acceleration of the radial (red) and lateral (blue) expansion of the cavity obtained by numerical differentiation of the data points (dots) and the smoothed curves (lines). The shaded regions outline the error ranges obtained from the smoothed curves.
(d) Aspect ratio (blue) and angular width (black) of the cavity (derived from measurements in \cite{long18}). }
	\label{cavity_kinematics}
\end{figure}

\newpage
\subsubsection{Early dynamics of the flux rope/cavity}

In the following, we study the dynamics of the flux rope/cavity by measuring its radial and lateral expansion and  outward motion. We use the measurements of \cite{long18}, who manually identified the inner edge of the bright rim around the cavity separately for four different AIA passbands, and then fitted these contours by an ellipse, extracting its minor and major axis as well as the height of its center above the limb (cf.\ Figure 2 therein). Here we use the width (minor axis) and height of the cavity center derived from the AIA 211 {\AA} images, as they contain the most complete  set of measurements. 
Note that these measurements of the kinematics of the dark cavity observed in AIA 211 {\AA} (inner edge)  basically correspond to the outer boundary of the bright rim in AIA 131 {\AA} (cf.\ Figure \ref{fig_ellipse} and 
Sect.~\ref{sect_morphology}).
In addition, we use the measurements of \cite{seaton18}, who visually tracked the center of the rising flux rope and cavity combining all SUVI filters to derive its height evolution (cf.\ Figure 2 therein). The additional value of the SUVI data is that we can follow the flux rope across a larger height range than in AIA, and thus measure its full impulsive acceleration phase.

In Figure \ref{cavity_kinematics}, we show the results for the kinematics and dynamics of the CME cavity/flux rope, using the same methods as for the analysis of the CME shell described in Sect. \ref{sect_shell}. 
We plot (a) the height of the center of the flux rope/cavity and its width, (b) the derived velocity and (c) acceleration profiles of the cavity height and width, as well as (d) its aspect ratio and angular width.
As can be seen in panel (a),  the measurements of the cavity/flux rope height from \cite{long18} and \cite{seaton18} are in agreement with each other despite being derived from different instruments and with different methods. 
The radial propagation of the cavity/flux rope starts with  a slow rise phase (first measurements of \cite{seaton18} are from 15:44 UT), changing to a faster increase around 15:51 UT. At  15:57:30 UT at a distance of 0.5~$R_s$ above the limb (when it exits the SUVI FOV), it reaches  $\sim$1600~km~s$^{-1}$, and the speed is still increasing.  
The width of the cavity first reveals a slow continuous expansion, which then quickly increases at 15:53 UT, reaching an expansion speed of $\sim$500~km~s$^{-1}$ about 2 min later when it exits the AIA FOV; the corresponding speed of its radial motion at this time is $\sim$800~km~s$^{-1}$. 

The acceleration of the lateral expansion (cavity width) reveals a maximum of $5.3\pm 1.1$~km~s$^{-2}$ at 15:53:40 UT, while the acceleration of the radial motion of its center reaches a peak of $6.7\pm 0.9$~km~s$^{-2}$ at 15:55:20 UT. Notably, the radial acceleration reveals a more continuous profile, constantly increasing over about 6 min until reaching its maximum, whereas the acceleration profile of the cavity width is more impulsive with a full duration of less than 3 min.
The  aspect ratio (angular width) reveals a slow increase (decrease) in the early phase followed by a fast decrease (increase) starting at 15:53:20 UT, during which the angular width of the cavity increases from 18 to $28^{\circ}$ over just 2 min, indicative of a strong lateral overexpansion of the cavity during this phase.

\begin{figure*}
\includegraphics[width=0.9\textwidth]{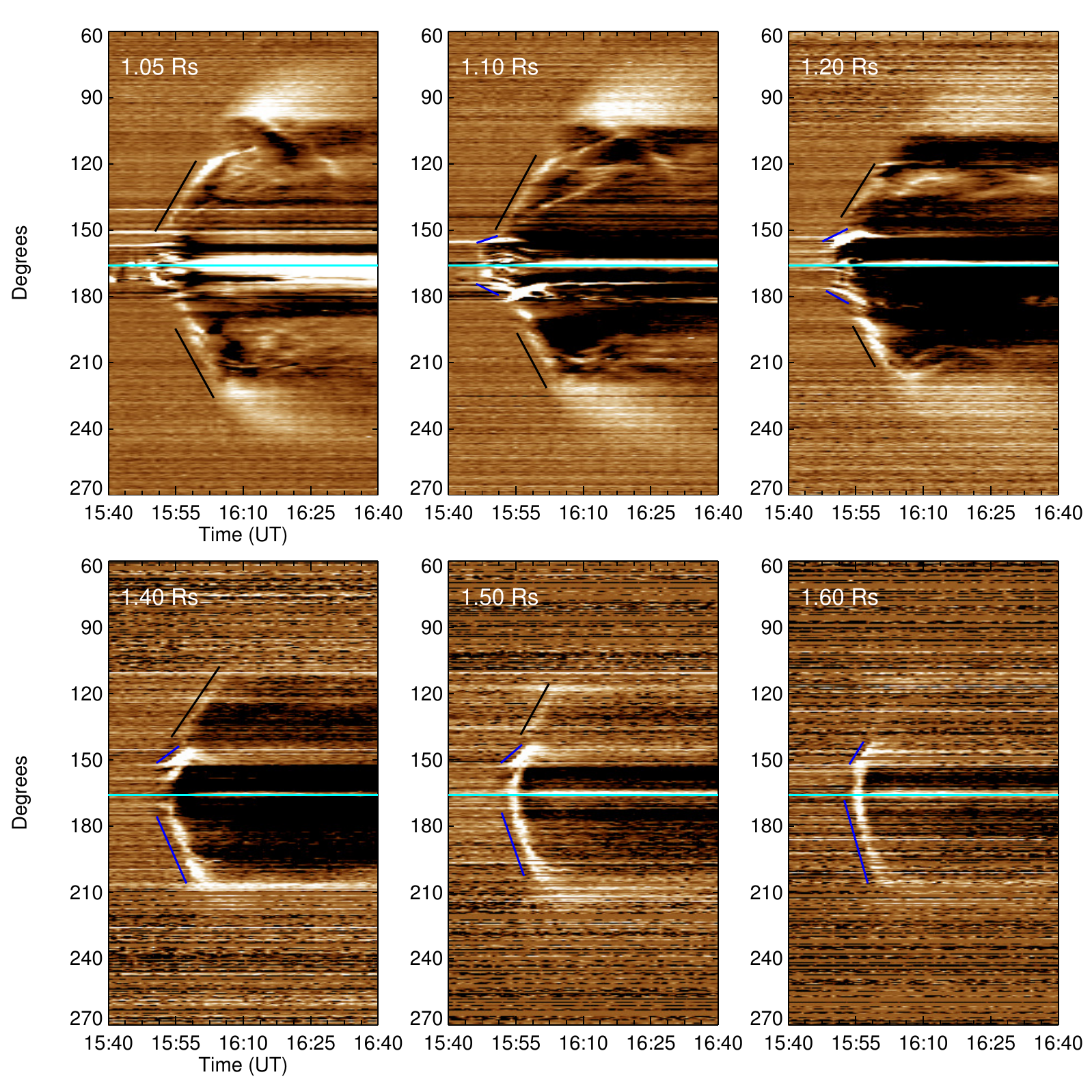}
\caption{Stack plots derived from SUVI base difference images in circular slits with increasing radius,  ranging from  $1.05$~$R_s$ to 1.6~$R_s$ (cf.\ Figure~\ref{fig_overview}) from Sun center.  The horizontal line marks the center of the eruption. The CME flanks and the EUV wave fronts are marked by blue and black lines, respectively. To keep the visibility of the tracks in the stack plots, the lines are shifted toward the left (by 2 min).}
\label{fig:suvi_stacks}
\end{figure*}

 \newpage
\subsection{Associated EUV wave}
\label{wave}

In order to study the formation of the associated EUV wave and its relation to the expanding CME shell,  we show in  Figure~\ref{fig:suvi_stacks} stack plots generated from SUVI {195 \AA} base difference images along circular slits at various distances from Sun center, corresponding to heights of 
$r=0.05$ to $r=0.6~R_s$ above the solar surface
(cf.\ Figure~\ref{fig_overview}, top right panel). The emission is averaged over a thin circular layer covering the width of 2 SUVI pixels (corresponding to 5 arcsec) across the considered radial distance, and then we stacked the slits obtained at different time steps. Each stack plot in Figure~\ref{fig:suvi_stacks} shows for a certain radial distance~$r$ from Sun center, the emission enhancement with respect to a pre-event base image as function of the polar angle $\varphi$ (on the $y$-axis) and time ($x$-axis). 
The counting of $\varphi$ starts at the left (eastern) horizontal line of the SUVI images, and the eruption center is located at $\varphi = 166^\circ$ (indicated by the cyan line in Figure~\ref{fig:suvi_stacks}).

The stack plots shown in Figure~\ref{fig:suvi_stacks} clearly show the impulsive lateral expansion of the CME flanks, the EUV wave that is formed and propagating ahead of the CME flanks, as well as the coronal dimmings behind. 
There is an obvious asymmetry in the behaviour toward the Northern and Southern directions. Toward the North, the EUV wave front is clearly detached from the expanding CME flank and propagating ahead of it. The EUV wave is formed as early as 15:52:24 UT at heights $\lesssim 0.2$~$R_s$ above the solar limb.
 This is 1.5 min before the peak of the acceleration of the lateral expansion of the CME shell, but at that time the lateral acceleration is already as high as  6~km~s$^{-2}$ (cf.\ Figure~\ref{fig3}).  
 Note that this formation time of the EUV wave  also well  coincides with the start of the strong decrease in the aspect ratio of the CME shell, i.e. the strong lateral overexpansion.  At the time of first EUV wave detection (15:52:24 UT), the distance of the wave front is only 38 Mm ahead of the contact surface, i.e.\ the lateral outer boundary of the CME shell. During its evolution, the distance of the wave front from the CME flanks quickly increases and the wave front  grows in height, reaching up to the edge of the SUVI FOV. In the stack plots in Figure~\ref{fig:suvi_stacks}, the track of the propagating wave front can be clearly identified and measured up to heights of at least 0.5 $R_s$ above the limb.
The behavior in the Southern direction is quite different. Here, it is only at times $\gtrsim$15:58~UT that the EUV wave can be distinguished from the CME structure, and it is observed only at lower heights, up to about 
0.3~$R_s$. 

In Figure~\ref{fig:suvi_stacks}, we have also indicated the outer fronts of the CME flank (blue lines) and the EUV wave (black lines), the latter being used to estimate the mean speed of the EUV front at different heights above the limb. The results reveal an increase of the EUV wave speed with height, in the Northern direction from  750~km~s$^{-1}$ at 0.05~$R_s$  to 1200~km~s$^{-1}$  at 0.5~$R_s$, with the strongest increase taking place between 0.4 and 0.5~$R_s$ above the solar limb.  In the southern direction, we also observe a height dependence of the EUV wave speed, increasing from 750 to 950~km~s$^{-1}$  from 0.05 to 0.3~$R_s$.

\section{Discussion and Summary}

The X8.2 flare/CME event of 10 September 2017 provides us with a unique opportunity to study the origin, morphology and  impulsive dynamics of  a very fast eruption ($v \gtrsim 3500$ km s$^{-1}$) as well as its relation to the fast EUV wave formed. In the present case we can observe and measure both, the evolution of the outer CME shell and the embedded flux rope/cavity. This is thanks to the combination of the high-cadence EUV imagery of SUVI and AIA, covering the low and middle corona, where the impulsive CME  dynamics takes place.  In addition, these unprecedented set of observations reveal important implications on the three-part structure of CMEs.

\subsection{Impulsive dynamics of the CME shell and cavity/flux rope}

The early evolution of the CME shell is well observed in the SUVI 195 {\AA} filtergrams. It develops from a set of very large pre-existing loops (with sizes up to $>$0.5~$R_s$). During their expansion, successively higher overlying loops are piled-up and become part of the erupting structure (cf.\ Figure \ref{fig_overview} and associated movie, and Figure \ref{fig2}). Comparison with the images by the LASCO C2 coronagraph shows that the outer front of these expanding and piled-up loops observed in the EUV images, seamlessly matches with the CME front observed in white-light coronagraph  images (Figure \ref{fig_lasco}).

The event also reveals a distinct cavity embedded in the CME shell, strongly indicative of the flux rope.
Magnetic reconfigurations indicative of reconnection are observed early in the event (most prominently in the AIA 131 {\AA} passband), connecting the forming flux rope structure to the Northern and Southern footpoints of the large-scale overlying loops observed in SUVI 195 {\AA} images. These reconnections appear to trigger the eruption, during which the cavity becomes well pronounced and  is quickly expanding and rising (cf.\ Figure \ref{fig_cavity} and accompanying movie).

The expanding cavity appears dark in all AIA channels, indicative of the density depletion inside the cavity due to the enhanced magnetic pressure of the enclosed flux rope \citep[e.g.][]{gibson15}, and it is surrounded by a bright rim.  The hot bright rim (most distinctly observed in AIA 131 {\AA} filtergrams) that develops around the expanding dark cavity has temperatures in the range $T \approx 10-15$~MK  \citep{long18,yan18}, and its lower boundary is connected to the rising cusp-shaped flare loop system by a hot ($T \approx 20$ MK) and thin elongated current sheet \citep{yan18,warren18}. In the cooler coronal AIA 171 {\AA} filter, reconnection inflows into the currents sheet were observed with velocities of about 100 km s$^{-1}$ \citep{yan18}. All these findings strongly suggest that the hot rim around the expanding cavity is a manifestation of the new flux and heated plasma that is added to the flux rope by magnetic reconnection in the current sheet beneath.
 
The role of reconnection for the CME dynamics lies in transferring magnetic flux from the ambient field into poloidal flux of the flux rope, which is  associated with the current in the flux rope.  This additional flux is important for the CME acceleration, as it strengthens the hoop force  acting on toroidal flux ropes \citep{shafranov66,chen89} and reduces the effect of the inductive decay of the electric current in the expanding flux rope \citep{vrsnak08b,vrsnak16b}. 
The reconnected flux typically changes with time from a considerably sheared to a nearly purely poloidal one, as indicated by the well known evolution of the flare loops in many events \citep[e.g., review by][]{fletcher11}. The ratio of toroidal and poloidal flux also differs from event to event. In the present case, the formation of a complete rim around the erupting flux rope indicates that the reconnected flux is dominantly poloidal from the beginning of the event (also see Sect.~\ref{sec_threepart}).

In the 10 September 2017 event, the CME shell and the cavity reveal an unprecedented fast evolution. The peak acceleration of both the lateral and radial expansion of the segmented CME shell are reached within the SUVI FOV. The lateral CME expansion reaches a peak acceleration of $10.1\pm 1.1$~km~s$^{-2}$, and in the radial direction we find a peak of $5.3\pm 1.4$~km~s$^{-2}$ at a height of 0.62 $R_s$ above the 
limb. Such high lateral acceleration values have not been reported before, and the radial acceleration also is among the highest values reported. Statistically, CME peak accelerations lie mostly in the range 0.1--1~km~s$^{-2}$ \citep{zhang06,vrsnak07,bein11,bak13}. 
To our knowledge, only six CMEs have been reported with peak radial accelerations of the CME front
$\gtrsim$5~km~s$^{-2}$ \citep{zhang01,williams05,vrsnak07,temmer08,bein11,ying18}. 
Lateral accelerations of the CME flanks are rarely derived, due to the difficulties in extracting the whole CME structure low in the corona. The highest values we found in the literature are in the range of 1--2~km~s$^{-2}$ \citep{patsourakos10b,patsourakos10,cheng12}.

We note that in a very recent paper on this event by \cite{gopalswamy18}, a peak value of the radial CME acceleration of $9.1\pm 1.6$~km~s$^{-2}$ at 15:58 UT at a height of 2.05 $R_s$ from Sun center is claimed.  However, looking into their kinematics plot (Figure 2), it is clear that this value is an artifact, arising from the transition between two instruments when changing the measurement data from EUVI to COR1, which results in a local jump of the distance-time curve. There is never a perfect transition in such combined measurements, because of different emission mechanisms and sensitivities of  different instruments. Thus, careful smoothing would be needed to properly account for such discontinuities. However, in  \cite{gopalswamy18} the velocity and acceleration profiles are derived from direct numerical differentiation of the height-time measurement points. As one can see from their Figure 2, the reported high (second) peak in the CME acceleration is obtained exactly at the time where the measurements change from EUVI to COR1, and is a mere artifact of the numerics when calculating the first and second time derivative without accounting for instrumental discontinuities and noise in the data.

The  acceleration of the width of the cavity occurs synchronized with the width of the CME shell. They reach their peak acceleration at 15:53:40 and 15:53:50 UT, respectively, with values of $5.3\pm 1.4$~km~s$^{-2}$ (cavity)  and $10.1\pm 1.1$~km~s$^{-2}$ (CME shell). The peak accelerations of their radial outward motion appear slightly later but again roughly synchronized, with a value of $6.7\pm 0.6$~km~s$^{-2}$ reached at 15:55:20 UT for the CME cavity and with a value of $5.3\pm 0.6$~km~s$^{-2}$ at 15:54:40 UT for the CME shell. These peak times of the radial accelerations are also close in time to the peak of the first RHESSI 100--300 keV HXR burst (15:54:40 UT) of the associated flare,  further underlining the importance of the reconnection process for the CME impulsive evolution.  The magnetic reconnection in the current sheet that is formed below the erupting flux rope provides for the energy release of the flare as well as additional magnetic flux to the flux rope to further facilitate its acceleration. This coupling and feedback relation establishes a synchronization of both processes, i.e.\ the flare energy release, associated particle acceleration and the CME dynamics \citep[][]{zhang01,neupert01,vrsnak04,temmer10,berkebile-stoiser12}.

Here, we demonstrate for the first time that the distinct hot rim observed around the flux rope cavity is a manifestation of the dominantly poloidal flux and frozen-in plasma that is subsequently added to the expanding and rising flux rope by reconnection in the strong event under study. An implication of these unique observations is that due to the very strong fields of up to 5~kG \citep{wang18} and the large system of densely packed loops in the source AR NOAA 12673, as observed in the EUV images, the values for the magnetic flux and plasma densities of the reconnnected loops are very high. 
These findings also  explain why in this event, the (lateral and radial) accelerations in the flux rope and CME shell reach such extremely high values;  all peak values lie in the range 5--10~km~s$^{-2}$. \cite{vrsnak16b} modeled a curved flux rope anchored in the photosphere, being subject to the kink and torus instability. He found that in the most strongly accelerated eruptions, in which peak values of the order of 10~km~s$^{-2}$ are reached, the poloidal flux supplied by magnetic reconnection has to be several times larger than the initial flux in the pre-erupting flux rope.

\subsection{Overexpansion and relation between CME shell and cavity evolution}

The CME shell and the cavity reveal not only an unprecedented fast evolution, but also a very strong overexpansion in the low-to-mid corona, i.e.\ the sizes (widths) of the structures expand much faster than they gain in height \cite[cf.][]{patsourakos10b,patsourakos10}. Further out in the coronagraphic FOV, generally an approximate self-similar CME evolution is observed, i.e.\ the CME size and height increase at the same rate \citep{schwenn05}, keeping the angular width and aspect ratio approximately constant. 
For the early evolution of the CME shell, we find an impulsive decrease of the aspect ratio from 0.98 to 0.75 during 15:53 to 15:57~UT, corresponding to  an increase of its angular width (measured from its source region on the limb) from about 90 to $105^{\circ}$. \cite{guo18} derived an angular width of about $130 ^{\circ}$ from Graduated Cylindrical Shell (GCS) fitting of the CME observed by the coronagraphs onboard LASCO and STEREO-A. This indicates that the CME width was still increasing after it  has left the SUVI FOV, which is consistent with the still increasing evolution seen in Figure~\ref{fig3}f.
For the cavity, we find first (from 15:50 to 15:53~UT) a slight decrease of the angular widths, before changing to an impulsive increase during 15:53 to 15:55~UT (when it exits the AIA FOV) from 18 to $28^{\circ}$ (Figure~\ref{fig_cavity}d).

It is important to note that, in lateral direction, the CME shell expands much faster than the cavity (compare the red curves in Figures~\ref{fig3}d and~\ref{cavity_kinematics}b). This is also true for the loops {\em between} the cavity (flux rope) and the CME shell (see the 171, 193, and 211~{\AA} channels in the animation of 
Figure~\ref{fig_cavity}). This means that, at these heights, these structures are not piled up by the expanding flux rope, but rather move away from it. More specifically, the CME shell is piled up by the expanding motion of ambient coronal loops in the intermediate environment of the flux rope (in a distance range comparable to the height of the flux rope) as the loops move away from the flux rope and press against the more distant coronal environment. This process obviously is part of the genesis of the CME, as it contributes to the formation of the CME shell and its enclosed cavity in the early stages of the eruption (roughly within $1~R_\mathrm{s}$ above the photosphere). The expanding motion of ambient flux, away from the rising flux rope, has been noted by \cite{patsourakos10b}, who suggested that it is a consequence of decreasing current through the flux rope as it rises. \cite{Kliem14} termed this a ``reverse pinch effect'' and confirmed it in MHD simulations of erupting flux ropes (full manuscript in preparation). 

\begin{figure}
\centering
\includegraphics[width=0.5\textwidth]{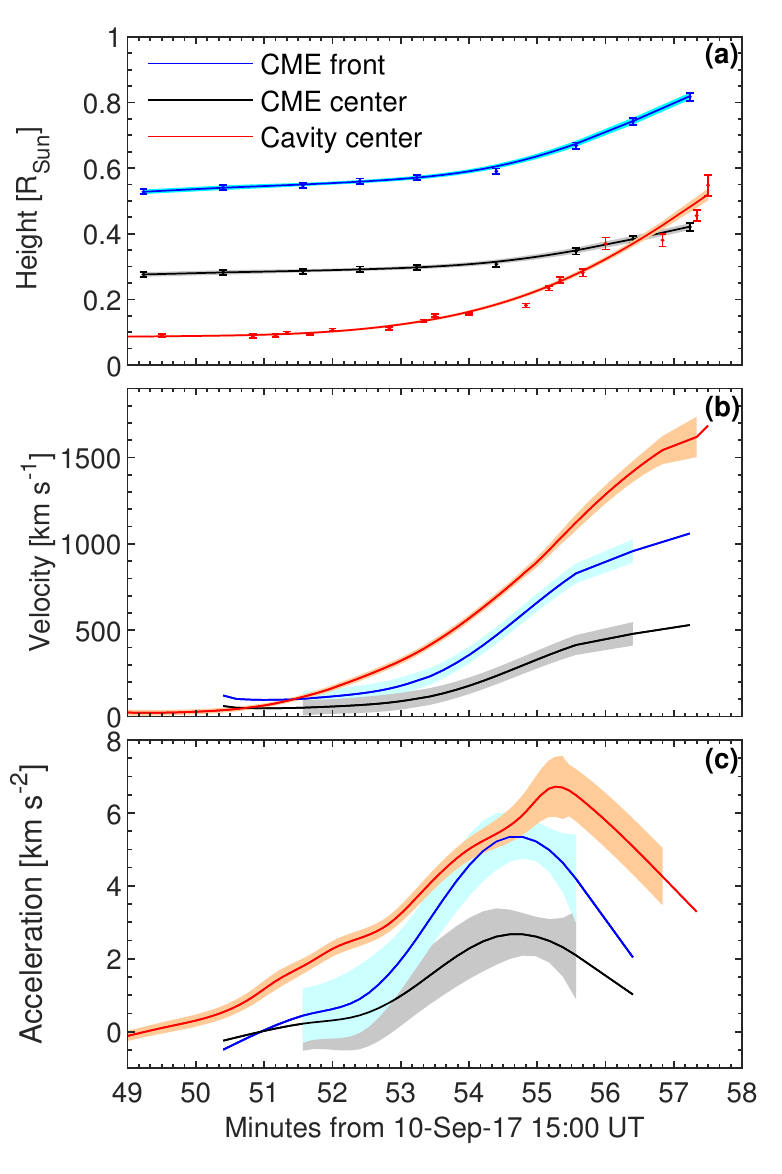}
	\caption{
	Combined radial evolution of CME shell and cavity.  
	(a) Height of the outer front of the CME shell (blue dots), the center of the CME shell (black dots) and the center of the cavity (red  dots) together with error bars. The corresponding lines show the smoothed height-time  profiles. (b) Corresponding velocity, and (c) acceleration profiles with the shaded regions outlining the error ranges obtained from the smoothed curves. }
	\label{kinematics_combined}
\end{figure}

In the radial direction, the situation is reversed: here the cavity (flux rope) rises faster than the overlying loops and CME shell (blue curves in Figures~\ref{fig3}d and~\ref{cavity_kinematics}b). This speed difference in the radial direction is seen more clearly in Figure \ref{kinematics_combined}, where we jointly plot the kinematics of the outer front of the CME shell, the center of the CME shell, and the center of the cavity.  It is clearly seen how the cavity moves forward inside the CME shell (panel a) from close to its bottom to beyond its center, due to its much higher velocity (panel b).
Since the reverse pinch effect should operate in the vertical direction in the same way as in the lateral direction, the different behavior, which results in the lateral overexpansion of the CME shell, can only be explained by a stronger restraining force of the ambient corona in the vertical direction. The preexisting set of prominent large-scale coronal loops which pass over the eruption site at about $0.5~R_\mathrm{s}$ may provide this through a primarily downward directed tension force. High-reaching overlying loops typically run nearly perpendicularly to the polarity inversion line, so that they resist the upward expansion of an erupting flux rope more than the lateral expansion, especially the lateral expansion perpendicular to the loops. Additionally, the perturbation of a dominantly gravitationally stratified ambient corona by a developing CME can lead to a preferentially downward restraining force and lateral overexpansion of the CME \citep{pagano13}. However, the plasma beta in an active region with very strong photospheric field strengths (up to 3 times higher than those assumed in \cite{gary01}) at heights of $\sim\!0.5~R_\mathrm{s}$ is still much smaller than unity, so that in the present event, the pressure gradient force of the ambient corona is less likely to cause the overexpansion than the strong overlying loops.

\subsection{Implications on the CME three-part structure}
\label{sec_threepart}
 
The distinct observations of the 10 September 2017 eruption provide us with important insight into the magnetic morphology of the erupting structure. We observe a hot bright rim around a quickly expanding dark cavity, which is embedded inside a much larger CME shell. The CME shell develops from a dense set of large pre-existing AR loops (with sizes up to $>$0.5~$R_s$). During their expansion, successively higher overlying loops are piled-up, and become part of the overall  erupting structure (cf.\ Figures \ref{fig_overview} and \ref{fig2}). The CME shell is best observed in the cooler filters of the SUVI and AIA  instruments (171, 195, 211~{\AA}) sampling plasma at quiet coronal temperatures around 1--2 MK. The hot rim around the expanding cavity is most distinctly observed in the AIA 131~{\AA}  filter, and has plasma temperatures in the range 10--15~MK \citep{long18,yan18}. It is a manifestation of the flux and frozen-in plasma that is added to the rising flux rope by magnetic reconnection in the current sheet beneath. As the plasma trapped on the coronal loops overlying the erupting structure is brought into the current sheet below by the reconnection inflow, it gets heated by the magnetic reconnection process. 
The heated plasma frozen to the downward closing flux builds up the hot flare loop system, whereas the heated plasma frozen to the upward ejected poloidal fields forms the expanding hot rim around the flux rope. 

The event under study also shows exceptionally well that the lower boundary of the hot rim around the cavity is connected to the rising flare loop system by a large-scale thin vertical current sheet, almost perfectly like in the eruptive flare cartoon in Figure~1 in \cite{lin04}. \cite{gary18} report sources of nonthermal microwave and hard X-ray emission indicative of accelerated electrons above the hot cusp-shaped flare loops, with the microwave sources extending even higher up into the indicated current sheet region. The thermal plasma in the elongated current sheet has been diagnosed to have temperatures of about 20~MK \citep{warren18}. These observational findings also provide strong evidence that in the present case the dark cavity outlines the cross-section of the flux rope, which is quickly expanding due to the continuously added  flux by the magnetic reconnection in the current sheet behind the erupting structure. 
We note that the formation of a bright rim around quickly expanding flux ropes due to the reconnected poloidal flux and plasma has been derived in the framework of the loss-of-equilibrium model \citep{lin04}. However, to our knowledge, this is the first event showing it in observational data in perfect agreement with the theoretical prediction.

What are the reasons that, in the event under study, all these structures are so well observed? ---In addition to the obviously favorable geometry with respect to the line of sight (i.e.\ that we look along the axis of the rising flux rope) and the excellent observational coverage of the low-to-mid corona by two wonderful EUV imagers, AIA and SUVI, that combine high-cadence, large FOV and multi-temperature imagery, there are the following effects. First, the pre-eruptive structure in the source AR is special. The AR has very high magnetic flux densities (up to 5 kG; \cite{wang18}), and contains of a set of dense bright loops reaching up to very large  heights ($\gtrsim$0.5$R_s$), as is indicated by their intense EUV emission in the temperature range 1--2 MK (cf.\ Figure \ref{fig_overview}). 
In the course of the eruption, this set of loops reconnects in the vertical current sheet. This means that high-density plasma is brought into the current sheet, forming the downward closing  flare loops as well as the bright rim around the flux rope. From a general perspective, we expext that the mass that is brought into the cavity rim increases with the magnetic field strength, because the rate at which field is brought into the reconnection region is proportional to the local Alfv\'en speed \citep{lin04}. High inflow speeds ($v \approx 100$ km s$^{-1}$) into the reconnecting current sheet behind the rising flux rope have been reported for the event under study \citep{yan18}. Another phenomenon that may come into play is that due to the very strong fields involved and due to the rapid expansion (and associated density depletion and adiabatic cooling), the cavity appears very dark. In addition, in the cooler AIA channels 
(where the bright rim around the cavity is not as distinct as in the hotter ones), the cavity yields a particularly high contrast against the bright set of surrounding loops which reaches high up into the corona.

As discussed in \cite{lin04}, instead of the full hot rim around the erupting flux rope, more often  U- or V-shaped structures are observed at the top of the current sheet, like in the 10\% of white-light CMEs that are classified as disconnection events, 
or also observed by EUV imagers low in the corona 
\citep[e.g.,][]{regnier11,liu13}. The observations of U- or V-shaped structures suggest the existence of an organized flux-rope like structure \citep[cf.,][]{vourlidas13}.
The more frequent appearance of U- or V-shaped structures instead of a complete rim 
may be related to more moderate injection rates of the poloidal flux by magnetic reconnection \citep{lin04}. Another reason may be that in some events the field lines are not sufficiently strongly coiled around the flux rope. Only in the case of strongly coiled, i.e., dominantly poloidal fields, the field lines below the flux rope that come to reconnection are winding close to the apex of the flux rope, resulting in a high poloidal flux component of the freshly reconnected field which would manifest itself in  a complete bright rim structure. In the more general case, the flux directly above the erupting flux rope (``strapping flux'') is also strongly sheared. When this flux comes to reconnection, it coils with a strong toroidal, i.e.\ shear, component around the erupting flux rope. This means that these fields manage to wind around the flux rope only far away from the apex, i.e. over the flux rope legs, resulting in V- or U-shaped structures of the hot plasma, when we observe along the axis of the flux rope. The formation of a complete rim in the present event suggests that the newly reconnected flux around the flux rope is dominantly poloidal from the beginning of the event.

\begin{figure*}
\includegraphics[width=1.0\textwidth]{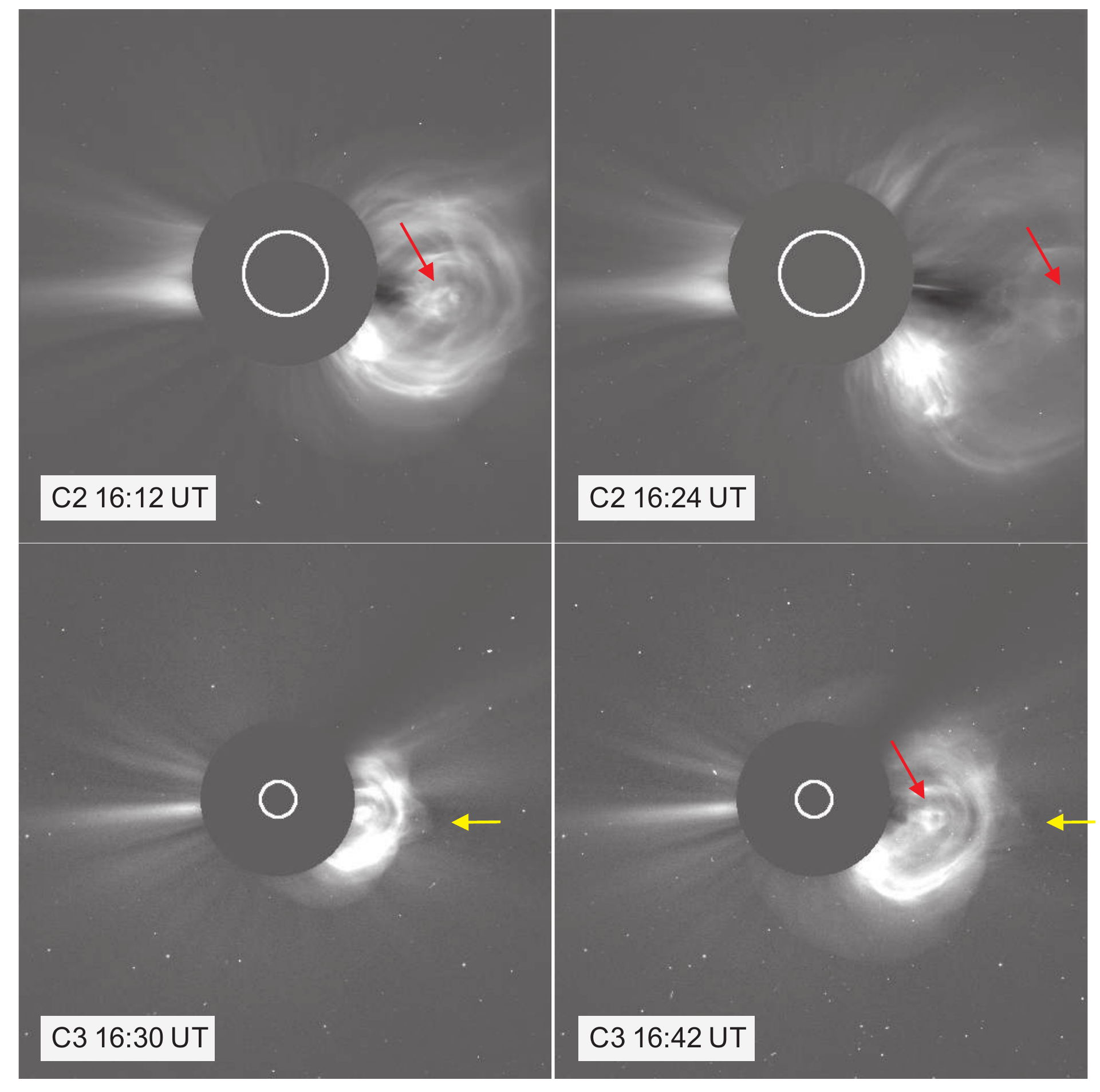}
\caption{  Sequence of LASCO C2 (top) and C3 (bottom) images of the CME revealing the extension of the bright rim around the cavity as observed in the EUV into the white-light coronograph data (indicated by red arrows). We also note a local deformation (``bulge''; indicated by yellow arrows) of the CME front/shock that is formed in the main direction of propagation of the  flux rope (core), as it moves much faster than the outer CME shell (cf.\ Figure~\ref{kinematics_combined}b). }
\label{fig_lasco2}
\end{figure*}

How do the structures of the eruption that we observe in the EUV connect to the traditional white-light coronagraph data? The composites of co-temporal SUVI/LASCO C2 difference images 
plotted in Figure~\ref{fig_lasco} clearly show that the expanding CME shell observed in the SUVI 195 {\AA} images fits to the CME front observed in the LASCO C2 data. In Figure~\ref{fig_lasco2} we plot a sequence of LASCO C2 and C3 direct images with less strong scaling to also obtain insight into the inner morphology of the white-light CME. In all four frames (as well as in later C3 images) we see a core with a bright rim-like structure embedded in the CME shell. In the C2 image at 16:12 UT, we can even see that this core has a tear-drop shape, very similar to the cavity/rim morphology observed in the EUV (cf.\ Figure~\ref{fig_cavity}). To make sure that this white-light structure corresponds to the cavity/rim observed in the EUV, we extrapolated the cavity height and width kinematics derived from the SUVI and AIA EUV images to the LASCO FOV.  
Estimating the distance of the center of the cavity from the radial kinematics curve in 
Figure \ref{cavity_kinematics}a (blue curve), we obtain a distance from the solar limb of $\approx$2.8 $R_s$ at 16:12~UT. This roughly fits with the distance of about 3.0~$R_s$ of the rim-like core observed in C2 at that time. We also cross-checked the correspondence of the width of  the hot rim observed in EUV and the width of the rim observed in the white-light coronagraph images. Applying a linear fit to the evolution of the cavity width in Figure~\ref{cavity_kinematics}a (red curve) to the data after 14:53:30 UT, we obtain a mean expansion rate of about 0.041~$R_s$ per minute. Extrapolating this to the time 16:12 UT, where the rim-like  structure is observed for the first time in LASCO C2, we obtain a width of 0.80~$R_s$. This  is consistent  with the width of the rim-like structure observed at that time in C2, which is about 0.8--0.9~$R_s$. Note that this correspondence also implies that in the LASCO FOV the overexpansion of the cavity has ceased and it is evolving almost self-similarly.

These findings imply that --- in the present event --- the core  of the CME observed in the white-light coronagraphs is {\em not} a manifestation of the dense prominence plasma, as is the traditional view of the three-part CME structure, but it is a manifestation of dominantly poloidal flux and hot plasma continuously added around the expanding flux rope by the magnetic reconnection in the current sheet beneath. This is also consistent with the observations that in this event there is only little evidence for erupting prominence plasma, which dissolves already very early in the eruption (see Figure~\ref{fig_cavity} and associated movie; cf.\ also  \cite{long18}). 
These findings strongly support the same conclusions in \cite{lin04} and \cite{howard17}, and have important implications. First, the bright CME core observed in white-light coronagraph images is not necessarily a signature of an embedded prominence. Second, the cross-section of the erupting flux rope can be significantly smaller than the CME cavity in white-light images, different from the widely accepted suggestion in \cite{chen97}.


Finally, we note that the white-light CME evolution in Figure~\ref{fig_lasco2} (most obviously in the C3 images) shows a distinct deformation (``bulge'') of the outer CME front. This deformation is  formed in the direction of motion of the expanding flux rope/cavity, which moves much faster than the outer CME shell, and is thus protruding from the bottom of the CME shell (cf.\ Figure~\ref{kinematics_combined}a,b). 
Such an accelerated motion of the flux rope with a speed considerably higher than that of the plasma overlying it, causes a continuous compression of the magnetoplasma ahead of the flux rope, inevitably creating an MHD wave signal that propagates through the overlying CME structures, including the sheath region in front of the eruption. Since the speed of the driver is increasing (cf.\ Figure~\ref{kinematics_combined}b, red curve), each new element of the wavefront has successively higher amplitude, implying that it propagates faster than the previous wavefront elements (for the relationship between the driver kinematics and the wavefront velocity profile and kinematics see \cite{vrsnak00}). Thus, the new wavefront elements sweep over the previous ones, overtaking them and heading towards the leading edge of the erupting structure. Finally, when the fastest elements reach the forehead of the structure, and continue the outward propagation faster than the flank elements of the erupting structure, the described effect causes a local deformation of the shape of the overlying large-scale CME shell and the associated shock front in the form of a bulge.
Such inhomogeneities and local regions of faster propagation speeds are a severe challenge and source of uncertainty in predictions of the arrival time/speed of the CME, its shock and the space weather effects it produces (see also \cite{torok18}).

\subsection{EUV wave formation}
 
The associated EUV wave reveals high speeds, increasing with height from about 750~km~s$^{-1}$ at 0.05~$R_s$  to 1200~km~s$^{-1}$ at 0.5~$R_s$ above the solar limb. 
Such a behaviour is also found in numerical simulations (e.g., Figure 6 in \cite{vrsnak16}), and it is most likely related to the combination of two effects: the ambient Alfv\'en speed increasing with height \citep{mann99} and the flux rope overexpansion (e.g., Figure 2c in \cite{vrsnak16}). 
Toward the Northern direction, the EUV wave can be distinguished from the driver (the CME flank) already at a distance as close as about 40 Mm at 15:52:24 UT. This is significantly smaller than  typical distances of 100--200 Mm at which EUV and Moreton waves become observable \citep{warmuth04a,kienreich11}.  As shown in the simulations in \cite{vrsnak16},  shock formation close to the contact surface implies a very 
impulsive acceleration of the driver. 
The early shock formation in this event is also supported by the associated metric type II burst, which starts at 15:53 UT \citep{gopalswamy18}.
A fast acceleration of the piston results in a higher wave amplitude and speed of the wave crest, and consequently to formation of the shock in a shorter time/distance. This is fully consistent with our observations of an extremely fast  lateral expansion of the CME with peak values up to 10~km~s$^{-2}$, and the formation of the EUV wave  close to the CME flank during the rise of its impulsive acceleration. Toward the Southern direction, the wave front can only later be distinguished from the CME  flanks, which implies either that the lateral expansion in this direction is not that impulsive as in the northern direction and/or that the ambient Alfv\'en speed is higher here, so the process of wavefront steepening is slower. 
 
The observations of the wave formation presented provide strong evidence for the interpretation of the global large-amplitude coronal EUV wave formation by a 3D piston mechanism (for the terminology and the physical background see, e.g., \cite{vrsnak05,warmuth15}). 
Firstly, the wave speed is much higher than the speed of the expanding CME flank. This is a typical signature of the wave formation by the piston mechanism (see e.g., the simulation results presented in Fig. 6b in \cite{vrsnak16}, showing that the wave is about two times faster than the piston), whereas in the case of a bow-shock scenario the speeds of the driver and the wave should be comparable 
\citep[][and references therein]{vrsnak05}. The higher speed of the wave as compared to the driver, indicative of a 3D piston mechanism, is well in line with the observations in the present case, where the CME flank expansion at different height levels is considerably slower than the EUV wave formed, by a factor 2--3 
(compare the blue and black lines outlining the CME flank and EUV wave propagation in the stack plots in Figure \ref{fig:suvi_stacks}).
Secondly, according to simulations presented in \cite{vrsnak16}, at low coronal heights (at the height of the flux rope center), the wave is formed and first appears during the impulsive lateral expansion of the magnetic structure, i.e.\ close in time to the CME peak acceleration and spatially close to its ``contact surface'' 
(see Fig. 6a,b therein, and the supporting observations in \cite{patsourakos10b} and \cite{cheng12}). This is in good correspondence with the distance/time we find in the observations of the event under study: 15:52:30 UT corresponds to the impulsive lateral acceleration phase of the CME shell, 
where the  lateral acceleration has already reached a value as high as 6~km~s$^{-2}$,  
and the impulsive decrease of the CME aspect ratio. Finally, we note that in the observations we identify a transient dimming travelling behind the wavefront (Figure \ref{fig_overview} and associated movie), which is expected in the case of wave formation due to the 2D and/or 3D piston mechanism \citep{landau87,vrsnak16}.
 
Interestingly, despite the strength, high speed and global propagation of this EUV wave \citep[see also][]{liu18}, it did not reveal any signature of an associated Moreton wave. This means that the pressure pulse initiated by the coronal wave was not strong enough to sufficiently perturb the underlying denser chromospheric plasma \cite[e.g.][]{vrsnak16}.  
This is probably due to the fact that in the present case the main lateral CME expansion --- although very strong and impulsive (with a peak acceleration of 10 km s$^{-2}$) --- happened at a comparatively large height and started from a large source region. Finally, we note that this event was associated with a wide-spread SEP event, covering $>$230$^\circ$ in helio-longitude \citep{guo18}. We may speculate that the formation of this wide-spread SEPs is also related to the extremely impulsive lateral expansion of the CME and the associated coronal shock formation close to the Sun.  

\section{Conclusions}

The CME associated with the X8.2 flare of 10 September 2017 provides us with unique observations on the genesis, the magnetic morphology and the impulsive dynamical evolution of a very fast solar eruption. 
In this event, the different parts of the CME can be clearly identified and followed in multi-temperature EUV images, and uniquely associated with the morphology in  the white-light coronagraph data. 

We clearly demonstrate that the hot bright rim observed around the quickly expanding EUV cavity is formed  by the dominantly poloidal flux and the attached heated plasma that is added to the erupting flux rope by magnetic reconnection in the large-scale current sheet beneath, and that this structure extends into the coronagraph FOV as CME  core. 
These findings corroborate the recently suggested rethink of the traditional view on the three-part CME morphology, which interprets the CME core as the signature of the erupting flux rope, rather than as the signature of prominence material trapped in the bottom part of the flux rope. Further studies are needed to quantify in what fraction of CMEs the core is a manifestation of the prominence and/or of the reconnected poloidal flux and plasma, and how we can use these observations to obtain estimates of the poloidal flux supplied to the eruption by magnetic reconnection and sustaining its acceleration. 
Our findings also imply that the cross-section of the erupting flux rope can be significantly smaller than the CME cavity observed in white-light images.

In the present event, the clearly defined CME structures together with the high-cadence large-FOV EUV imagery also allowed us to derive the impulsive dynamics separately for the radial and lateral expansions of both the CME shell and the flux rope, all of which revealed extreme acceleration values with peaks in the  range 5--10 km s$^{-1}$. The much higher radial propagation speed of the flux rope as compared to the CME shell is identified to be the reason for the substantial deformation of the white light CME front and shock in the form of a faster moving bulge. Such local inhomogeneities have important implications for space weather forecasts, as they pose additional difficulties in the prediction of the CME arrival time and speed at Earth. Finally, we note that both the CME shell and the flux rope exhibit a strong lateral overexpansion during the impulsive phase of the event. The overexpansion and the 2D/3D-piston mechanism are essential for generating the fast and globally propagating EUV wave associated.

\begin{acknowledgements}
A.M.V., K.D. and M.T. acknowledge the Austrian Science Fund (FWF): P24092-N16, P27292-N20, and the Austrian Space Appplications Programme of the Austrian Research Promotion Agency FFG (ASAP-11 4900217).
B.V. acknowledges financial support by the Croatian Science Foundation under the project 6212 {\it Solar and Stellar Variability}. B.K. acknowledges support by the DFG. D.M.L. received support from the European Commission's H2020 Programme under the following Grant Agreements: GREST (no.\ 653982) and Pre-EST (no.\ 739500).
\end{acknowledgements}

\begin{facilities}
GOES, RHESSI, SOHO
\end{facilities}


\end{document}